\documentclass[draft]{agujournal2019}
\usepackage{url} %this package should fix any errors with URLs in refs.
\usepackage{lineno}
\usepackage[inline]{trackchanges} %for better track changes. finalnew option will compile document with changes incorporated.
\usepackage{soul}
\let\linenumbers\relax

\draftfalse

\journalname{Earth and Space Science}

\begin{document}

\title{Ground-to-Cable Strain Transfer in Unburied DAS on Earth and the Moon}

\authors{S. Probst\affil{1}, O. Cornelius\affil{1}, M. Serra Garcia\affil{2}, J. O. A. Robertsson\affil{1}, C. M. Donahue\affil{3}}

\affiliation{1}{Department of Earth and Planetary Sciences, ETH Zurich, Zurich, Switzerland}
\affiliation{2}{AMOLF, Amsterdam, The Netherlands}
\affiliation{3}{Earth and Environmental Sciences, Los Alamos National Laboratory, NM, USA}

\correspondingauthor{Carly M. Donahue}{cmd@lanl.gov}

\begin{keypoints}
\item Unburied fiber-optic cables can lose signal when suspended cable segments absorb ground strain by bending instead of stretching.
\item Strain transfer depends only on the ratio of segment sag to cable radius and stays nearly complete below one quarter of the radius.
\item The sag comes from curvature retained from spooling rather than from gravity so the same criterion applies on the Moon.
\end{keypoints}

\begin{abstract}
Distributed acoustic sensing (DAS) cables are typically buried to ensure good coupling and reduce unwanted atmospheric noise. Unburied deployments are attractive for rapid response applications and for missions to the Moon, where burial is impractical. Unburied DAS, however, suffers from degraded signal quality due to poor strain transfer.
Here we quantify bending stress relief as a mechanism behind this loss: cable segments suspended between ground contact points accommodate ground strain by bending, so less axial strain reaches the fiber.
We derive the strain transfer of such a segment analytically, compare it to numerical models, and validate it on 56 laboratory configurations of different cable types and lengths.
The efficiency is governed by a single dimensionless parameter $\Theta$, set by the ratio of the segment's sag to the cable radius. Keeping the loss below 3\% requires a sag below about a quarter of the cable radius. The prediction is quasi-static, holding below the fundamental resonance of the segments.
The sag is dominated by the curvature the cable retains from spooling and handling rather than by gravitational sagging, so $\Theta$ has to be measured and not computed from cable specifications.
Because the gravitational contribution is small, the criterion applies equally to deployment on the Moon. Lunar gravity does reduce the friction available to resist slip at the contact points, but a force balance indicates that slip remains unlikely for natural moonquakes.
\end{abstract}

\section*{Plain Language Summary}
Fiber-optic cables can be used as seismic instruments. Stretching or compressing a cable slightly changes the light traveling through it, such that one cable can measure ground motion at thousands of points along its length. For this to work the cable has to move with the ground, which usually means burying it in a trench. Burial is costly, and impractical on the Moon, so there is interest in laying cables directly on the surface. Surface cables are known to record weaker signals, but the reason has not been fully understood.
We show that part of the loss comes from bending. A cable laid on uneven ground touches down only at discrete points and hangs freely in between. When the ground stretches, a hanging piece partly straightens out instead of stretching, and the fiber records less.
Whether this happens depends on a single metric: how much the hanging piece sags relative to the thickness of the cable. Measurements on seven cables, each at four lengths, confirm this. The sag comes mostly from the bend a cable keeps after being unrolled from its spool and not from its own weight, so the same rule applies under the weaker gravity of the Moon.

\section{Introduction}
\label{sec:introduction}
DAS is a fiber-optic sensing technique that measures the dynamic axial strain along a cable by analyzing phase shifts in Rayleigh-backscattered laser light \cite<e.g.,>[]{hartog_introduction_2017, lindsey_fiber-optic_2021}. 
DAS records the elongation or shortening of an optical fiber, leading to an inherently one-dimensional measurement of strain in line with the sensing cable. 
Over the past decade, DAS has been adopted widely for borehole seismic applications such as vertical seismic profiling \cite{daley_field_2013,mateeva_distributed_2014}, and increasingly for surface deployments in which the cable is placed in a shallow trench and coupled to the ground by burial and compaction \cite{abukrat_applications_2023,hubbard_quantifying_2022}.
Burial serves two purposes: it shields the cable from wind-induced noise and ensures continuous mechanical contact with the ground, so that the strain is transferred from the surrounding medium onto the cable. 

Empirical results widely show that unburied DAS deployments, lacking this continuous contact, exhibit degraded signal quality relative to buried configurations. \citeA{an_traffic_2023} demonstrated this by comparing DAS recordings of traffic-induced vibrations on a road using the same cable in two configurations, cemented to the surface and surface-draped, finding that the latter recorded signals with significantly weaker amplitudes and lower signal-to-noise ratio (SNR). Similarly, \citeA{harmon_surface_2022} compared surface coupling strategies in a grassy field and found that a surface-draped cable could only capture hammer-source energy within approximately 10~m, while weighted configurations more than doubled this range. \citeA{zandanel_earthquake_2026} compared buried and unburied deployments in a lunar regolith simulant and found that unburied cables were able to record earthquakes but with lower amplitude compared to the buried configuration.
Part of this signal degradation can be attributed to atmospheric noise: wind exerts mechanical forces on the cable, contaminating the seismic signal and lowering the SNR \cite{hudson_distributed_2021,probst_controlled_2026,viens_rapid_2025}. However, poor signal quality persists even in wind-free environments: \citeA{probst_controlled_2026} showed in controlled laboratory experiments that unburied cables recorded lower amplitudes than buried cables even without any fan-induced wind, and that this loss was exacerbated where surface undulations prevented continuous contact with the ground. These observations indicate that poor ground coupling, and not atmospheric noise alone, is a fundamental challenge of unburied DAS.

Despite this challenge, unburied deployments are attractive, and in some settings indispensable.
Burial is costly, time-consuming and requires excavation infrastructure that is unavailable or impractical in many environments. This makes unburied deployments a necessity, for example, in rapid-deployment scenarios such as recording aftershocks following a large seismic event. 
Burial is also infeasible in the challenging lunar environment, where the fine-grained and abrasive regolith poses large risks to infrastructure. DAS has nonetheless been proposed for deployment on the Moon because its robustness and dense spatial sampling make it an attractive alternative to single-station seismometers. On the Moon, a rover could unroll kilometers of fiber-optic cable directly onto the surface, and record moonquakes, meteorite impacts, and active sources \cite{wu_fiber_2024,zhai_assessing_2024}. 
An unburied cable is held against the ground only by its own weight, which on the Moon is about one-sixth of its weight on Earth. Predicting the coupling of a lunar deployment thus requires a description of the strain transfer in which the role of gravity is explicit. The cost of such a mission leaves no opportunity to test coupling strategies or adjust the deployment in the field. The mechanisms governing ground-to-cable strain transfer must therefore be established beforehand.

A cable resting on or buried in the ground does not fully transfer ground strain to the fiber core. Two steps can each attenuate it: the transfer from the ground to the cable's outer surface, which we term ground-to-cable transfer, and the transfer from that surface inward through the cable's layers to the fiber core, which we term cable-to-fiber transfer. The latter is the better characterized of the two, and we summarize it first before turning to ground-to-cable transfer, the focus of this study.
Cable-to-fiber transfer is governed by shear-lag mechanics across the concentric layers of the cable's cross-section (jacket, buffer, coating, cladding, and fiber core). Compliant intermediate layers attenuate the strain before it reaches the fiber core, as in loose-tube telecommunications cables, where the fiber is only weakly coupled to the jacket \cite{falcetelli_strain_2020,hubbard_quantifying_2022,reinsch_mechanical_2017,tan_strain_2021}. Empirical comparisons confirm this by consistently showing that tight-buffered cables yield higher strain transfer rates than loose-tube cables \cite{castongia_experimental_2017,forbriger_calibration_2025}.

For buried deployments, ground-to-cable transfer has been studied analytically and numerically. \citeA{zhang_toward_2020} developed a theoretical framework for the ground-backfill-cable interaction for distributed strain sensing in boreholes. \citeA{celli_full-waveform_2023} studied cable-ground coupling for DAS using a spring model, finding that the stiffness of the material immediately surrounding the cable is the dominant factor, leading to amplitude amplification and phase delays. 
For buried DAS, the physical mechanisms of both steps are reasonably well characterized. This study concerns unburied cables, for which the first step in this chain, the ground-to-cable transfer, remains poorly understood.
Contact is intermittent as the cable bridges over surface irregularities and touches the ground only at discrete points. The tangential force available to transmit axial seismic strain from the ground to the cable is limited by Coulomb friction at these contact points.
Existing research on unburied DAS deployments is almost entirely empirical. Field experiments have shown that adding weights such as sandbags improves coupling \cite{harmon_surface_2022}, though not consistently \cite{forbriger_calibration_2025}. Pressing a cable into snow improved coupling compared to draped deployment \cite{mjehovich_rapid_2023}. 

\citeA{probst_controlled_2026} found that thicker and stiffer cables perform better in unburied configurations and proposed bending stress relief as a key mechanism: a cable spanning contact points tends to bend rather than deform axially, so that the ground-induced axial strain is partially absorbed by cable bending instead of reaching the fiber as elongation and contraction. Thicker, stiffer cables resist this bending more effectively and could therefore record ground strain more reliably. However, this mechanism remained a qualitative hypothesis: it had not been tested or linked analytically to cable parameters.   

A second potential mechanism reducing the strain transfer in unburied deployments is frictional slip between the cable and the ground at the contact points. With nothing holding it down but its own weight, an unburied DAS cable depends entirely on frictional contact for its coupling \cite{celli_full-waveform_2023}. If the amplitude of the ground-induced axial force exceeds the maximum static friction available at the contact points, the cable will slip and reduce the axial strain transmitted to the fiber. 
Both mechanisms, bending stress relief and frictional slip, are expected to depend on cable parameters including the bending stiffness $EI$ (product of Young's modulus $E$ and area moment of inertia $I$), axial stiffness $EA$ (product of Young's modulus $E$ and cross-section area $A$), mass per unit length $m = \rho A$, outer radius $r$, and the friction coefficient $\mu$ between the cable and the surface. 
Both mechanisms also depend on gravity, which sets the normal force at the contact points and influences the shape of the deployed cable. 
At present, neither mechanism has been established as the physical explanation for the observed coupling losses, and to our knowledge no model exists that predicts how well a given cable couples to the ground in a surface deployment. 

The goal of this paper is to address this gap by identifying and quantifying the physical mechanism governing ground-to-cable strain transfer in unburied DAS deployments.
We model surface-draped cable segments as beams suspended between two contact points and derive the axial strain transferred from the ground to the cable.
This yields a dimensionless number, which predicts the strain transfer efficiency of a given segment.
We test the prediction against finite element simulations and against laboratory measurements on 56 suspended-segment configurations of seven cables, in which the segment geometry and the transferred strain were measured directly.
The measurements further show that the sag of a draped cable is set by the curvature it retains from storage and not by its own weight.
Our results provide a quantitative framework for understanding and optimizing unburied DAS coupling, with implications for cable design on Earth and on the Moon.

\section{Analytical Strain Transfer Model}

\subsection{Treatment of Cable as Segments Suspended Between Ground Contact Points}

\begin{figure}[!tb]
    \centering
    \includegraphics[width=\textwidth]{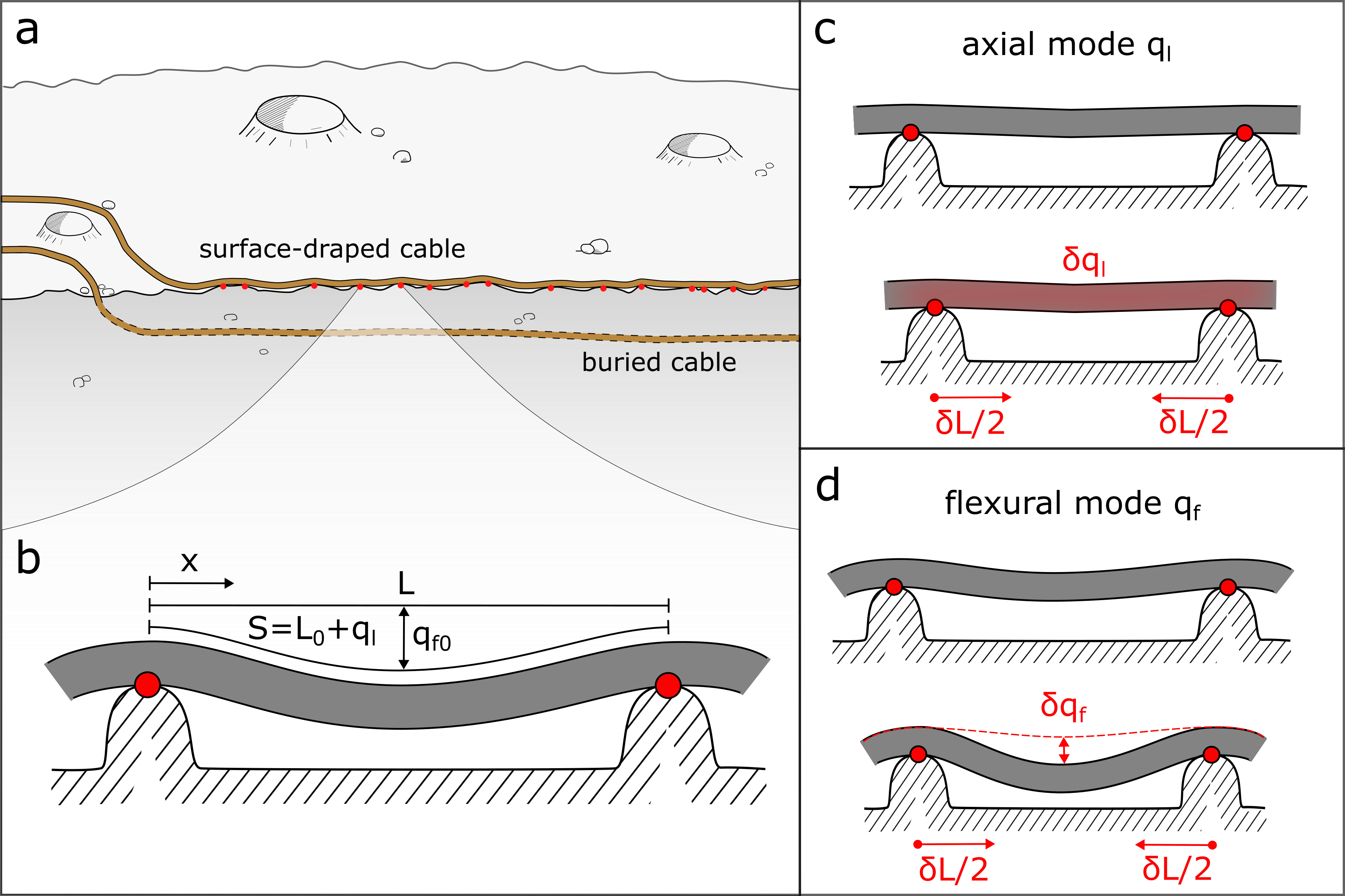}
    \caption{\textbf{Unburied cables lack continuous ground contact, and form suspended segments between discrete contact points that accommodate applied ground strain through a combination of axial and flexural deformation modes.} 
    \textbf{(a)} Illustrative setup of a fiber-optic cable surface deployment in a lunar environment. 
    \textbf{(b)} Zoom-in on a single suspended segment showing the variables used in the analytical derivations: the horizontal span between endpoints $L$, the arc length $S$, the unstretched arc length $L_0$, the initial midspan sag $q_{f0}$, and the axial deformation $q_l$ (change in arc length). 
    The cable segment has an initial bend, which may originate from curvature retained by the cable or be produced by gravity.
    When the segment endpoints are deformed by an applied ground displacement $\delta L$, this strain is partitioned into two principal deformation modes. 
    \textbf{(c)} The axial mode, where the cable accommodates the displacement by physically changing its arc length. The axial strain from the ground is transferred to the cable and can be measured by DAS. 
    \textbf{(d)} The flexural mode, where the cable bends to accommodate the displacement without changing its actual arc length. The axial strain from the ground is not transferred to the cable and cannot be measured by DAS.}
    \label{fig:coupling_and_modes}
\end{figure}

An unburied, surface-draped cable is modeled as a sequence of suspended segments between contact points (Figure~\ref{fig:coupling_and_modes}a). The spacing between ground-cable contact points depends on the roughness of the surface, on the cable's stiffness and radius, and on gravity, which sets its weight. Together, these determine how closely the cable conforms to the surface: a heavy, compliant cable follows small undulations, creating closely spaced contact points, whereas a stiff or light cable bridges over them and makes contact less often. This characteristic segment length would have to be measured for a given deployment and is treated here as an input to the model. 

When strained, hanging segments can bend instead of deforming axially. How the strain is transferred from the ground via the contact points to the cable itself is calculated using beam theory. The hanging cable segments are modeled as geometrically nonlinear Euler-Bernoulli beams that are deformed by the movement of their endpoints, the contact points with the ground. The following derivations and the numerical implementation quantify a single suspended segment. The recorded DAS response is the average over the gauge length, which spans many such segments of varying length. We focus here on characterizing the behavior of an individual segment, which is the building block of this averaged response. 
We consider the cable segment as a horizontal beam (Figure~\ref{fig:coupling_and_modes}b) with mechanical properties: unstretched arc length $L_0$, radius $r$, cross-sectional area $A=\pi r^2$, moment of inertia $I = \pi r^4 / 4$, Young's modulus $E$, and density $\rho$.
Here $L_0$ denotes the material length of the segment, that is, the arc length it takes up when it carries no axial tension.

To allow for an analytic treatment of the problem, the following assumptions are made.
First, the segment is assumed to carry an initial bend. It may arise from the cable sagging under its own weight, from curvature retained by the cable itself, for example after storage on a spool, or from a combination of the two. The derivation in Section~\ref{subsec:analytical_solution} is independent of the origin of the sag and treats the resulting midspan sag $q_{f0}$ as a given property of the segment. Section~\ref{subsec:what_sets_sag} then discusses the origin and magnitudes of the sag.
Second, this initial bend is assumed to be symmetric about the midspan and single-lobed, so that it is well approximated by a single flexural mode shape. Bends with an antisymmetric, S-shaped component are not captured by this description.
Third, the treatment is limited to an in-line strain field with wavelengths much larger than the cable segment, such that the strain is essentially constant over the length of the segment. Fourth, the segment is assumed to have no initial axial tension. Finally, no slip or rotation at the contact points is assumed: the motion of the cable at the contact points is identical to the ground motion at these points. 
This also implies no interaction between adjacent suspended segments, since the behavior of each segment is fully dictated by the movement of its endpoints.

The hanging cable segment at rest is thus modeled as a beam with doubly-clamped endpoints with an initial bend of midspan amplitude $q_{f0}$ (Figure~\ref{fig:coupling_and_modes}b). When the endpoints are displaced, the cable deforms to accommodate the applied displacement. The deformation is limited to two modes for this simplified treatment. 
The two deformation modes are axial elongation $q_l$ and the first flexural bending mode $q_f$ (Figure~\ref{fig:coupling_and_modes}c and d). The axial mode $q_l$ is the change in the beam's arc length, while $q_f$ is the maximum transverse deflection from the straight beam, which occurs at the midspan. In the initial rest state, $q_f$ equals the initial sag $q_{f0}$.
The transverse deflection is approximated as $w(x) = q_f \phi(x)$, where the first symmetric mode for a doubly-clamped beam is chosen as 
\begin{equation}
\phi(x) = \frac{1 - \cos\left(\frac{2\pi x}{L}\right)}{2},
\label{eq:mode_shape}
\end{equation} 
where $L$ is the horizontal span between the endpoints.
This satisfies the boundary conditions \(w(0) = w(L) = 0\) (beam endpoints fixed), and \(w'(0) = w'(L) = 0\) (beam endpoints orientation fixed), and gives $w(L/2) = q_f$ at the midspan. The mode shape assumes clamped rather than pinned endpoints: it prescribes both the deflection and the slope at the contact points. 
Clamping is the appropriate condition because the cable cannot pass through the surface, so at a contact point it must be tangent to it. The slope there is therefore set by the local ground surface. This argument applies to a cable bent in the vertical plane. A bend in the horizontal plane is not constrained in the same way, since friction fixes the position of the contact point but not the tangent. The consequences of adopting pinned conditions instead are discussed in Section~\ref{subsec:limitations}.

DAS primarily measures changes in optical path length along the fiber. In the present treatment, we assume that the DAS response is dominated by axial elongation of the cable segment and we neglect potential effects of fiber curvature on the optical measurement. Under this assumption, only the axial deformation mode contributes to the measured DAS signal while strain accommodated by bending is not recorded.
The static strain transfer efficiency describes the fraction of the axial strain which is transferred from ground to cable and is defined as $\frac{\delta q_l}{\delta L}$, where $\delta q_l$ is the axial length change of the cable segment and $\delta L$ is the change of the horizontal span.  

\subsection{Analytical Strain Transfer Efficiency}
\label{subsec:analytical_solution}

The system we consider starts from an unperturbed rest configuration, in which the suspended segment carries an initial bend of midspan amplitude $q_{f0}$. 
When a seismic wave arrives, it displaces the segment's endpoints slightly, perturbing this state. The segment responds by partitioning the displacement between its two deformation modes, bending and axial elongation. 
We assume that only the axial component is transferred to the fiber as measurable strain, so this split between modes is what determines the strain transfer efficiency.
To determine the transfer efficiency, we first establish the geometric constraint that couples the two modes, then we linearize the geometric constraint about the rest state and minimize the incremental energy to calculate how a small endpoint displacement is partitioned between them.

The arc length of the beam is given by $S = \int_0^L \sqrt{1 + (w'(x))^2} \,dx$ \cite<e.g.,>[]{reddy_energy_2002}.
Assuming small slopes $(|w'(x)|\ll1)$, expand to second order:
\begin{equation}
S\approx \int_0^L \left( 1 + \frac{1}{2} (w'(x))^2 \right) dx = L + \frac{1}{2} \int_0^L (w'(x))^2 dx.
\end{equation}
Substitute $(w(x) = q_f \phi(x))$:
\begin{equation}
S = L + \frac{1}{2} C_1 q_f^2, \quad C_1 = \int_0^L (\phi'(x))^2 dx.
\end{equation}
The excess of arc length over span, $\frac{1}{2}C_1 q_f^2$, is the length stored in the bend. A segment can therefore change its span without changing its arc length, by changing how much of its length is stored this way. 
The arc length of the beam must equal the unstretched arc length plus axial elongation $S = L_0 + q_l$. Thus, bending reduces the effective horizontal span for a given axial length: 
\begin{equation}
\label{eq:geometric_constraint}
L = L_0 + q_l - \frac{1}{2} C_1 q_f^2.    
\end{equation}
This geometric constraint is the only place where we distinguish between the unstretched arc length, $L_0$, and the horizontal span, $L$; elsewhere, for the evaluation of the coefficients $C_1$ and $C_2$ and the energies below, we set $L_0 \approx L$ using the same small-slope assumption already used to expand the arc length. 
We consider a small ground movement $\delta L$, which causes small changes
$q_f = q_{f0} + \delta q_f$ and $q_l = \delta q_l$. Linearizing about the
initial state ($q_l=0$, $q_f=q_{f0}$):
\begin{equation}
\label{eq:linearized_constraint}
\delta L = \delta q_l - C_1 q_{f0}\, \delta q_f.
\end{equation}

We next require the energy cost of the perturbation, measured from the rest state. The total potential energy of the segment is $\Pi = U_l + U_f + V$, where
\begin{equation}
\label{eq:energies}
U_l = \frac{1}{2}\frac{EA}{L} q_l^2, \qquad
U_f = \frac{1}{2} EI\, C_2 \left(q_f - q_{f0}^{\mathrm{bend}}\right)^2, \qquad
C_2 = \int_0^L (\phi''(x))^2 dx ,
\end{equation}
assuming small strains, so that the cable is governed by the equations of linear elasticity and the axial stiffness $EA/L$ may be treated as constant \cite<e.g.,>[]{weaver_vibration_1990}. Strain energy is measured from the stress-free configuration, so $q_{f0}^{\mathrm{bend}}$ is the amplitude of any curvature the cable retains when unloaded (Section~\ref{subsec:what_sets_sag}); it is zero for a cable that relaxes to a straight shape. The term $V$ is the potential of any external load acting on the segment. For a static load such as gravity, it is linear in the deflection, $V_g = -\frac{1}{2} F_g q_f$, where $F_g$ is the gravitational force, and therefore contributes to the first-order terms of the expansion below but not to the second-order ones.
Expanding $\Pi$ about the rest state $(q_l, q_f) = (0, q_{f0})$ to second order gives
\begin{equation}
\label{eq:energy_expansion}
\delta \Pi = \left(\frac{\partial \Pi}{\partial q_l}\right)_{\!0} \delta q_l
+ \left(\frac{\partial \Pi}{\partial q_f}\right)_{\!0} \delta q_f
+ \frac{1}{2}\frac{EA}{L}(\delta q_l)^2
+ \frac{1}{2} EI\, C_2 (\delta q_f)^2 .
\end{equation}
No cross term in $\delta q_l \delta q_f$ appears, because $U_l$ depends only on $q_l$ while $U_f$ and $V$ depend only on $q_f$: the two modes are uncoupled in the energy, and all of their coupling resides in the geometric constraint (Equation~\ref{eq:linearized_constraint}). The flexural coefficient is $EI\,C_2$ irrespective of $q_{f0}^{\mathrm{bend}}$, since differentiating a quadratic twice is insensitive to a shift of its origin, so an initial bend neither stiffens nor softens the incremental flexural response at this order. 
Both first-order terms vanish. The first because the segment carries no initial tension, so $q_l = 0$ at rest. The second because the rest state is an equilibrium, $(\partial \Pi/\partial q_f)_0 = EI\,C_2 (q_{f0} - q_{f0}^{\mathrm{bend}}) - \frac{1}{2}F_g = 0$, which is the condition that determines $q_{f0}$ and is examined in Section~\ref{subsec:what_sets_sag}. The incremental energy is therefore
\begin{equation}
\label{eq:incremental_energy}
\delta U = \frac{1}{2} \frac{EA}{L} (\delta q_l)^2 + \frac{1}{2} EI\, C_2\, (\delta q_f)^2 ,
\end{equation}
independently of what produced the initial bend.

We minimize the incremental energy, $\delta U$, required to accommodate the ground movement, $\delta L$. 
Substituting the expression for $\delta q_l$ from the linearized constraint (Equation~\ref{eq:linearized_constraint}) gives: 
\begin{equation}
\delta U = \frac{1}{2} \frac{EA}{L} (\delta L + C_1 q_{f0} \delta q_f)^2 + \frac{1}{2} EI C_2 (\delta q_f)^2.  
\end{equation}
We minimize $\delta U$ with respect to $\delta q_f$ and solve for $\delta q_f$:
\begin{equation}
\delta q_f = - \frac{\frac{EA}{L} C_1 q_{f0} \delta L}{EI C_2 + \frac{EA}{L} (C_1 q_{f0})^2}.
\end{equation}
The negative sign expresses the expected behavior: moving the endpoints together drives the segment to bend further, while pulling them apart straightens it.
Substituting $\delta q_f$ back into the constraint gives 
\begin{equation}
\delta q_l = \delta L + C_1 q_{f0} \delta q_f = \delta L \frac{EI C_2}{EI C_2 + \frac{EA}{L} (C_1 q_{f0})^2}.    
\end{equation}
This yields the axial strain transfer efficiency,
\begin{equation}
\label{eq:strain_transfer}
\eta = \frac{\delta q_l}{\delta L} = \frac{1}{1 + \Theta}, 
\end{equation}
with 
\begin{equation}
\label{eq:theta}
\Theta = \frac{EA}{L} \frac{(C_1 q_{f0})^2}{EI C_2}. 
\end{equation}
Equation~\ref{eq:theta} holds for any admissible choice of mode shape $\phi(x)$, which enters only through the two integrals $C_1$ and $C_2$. Physically, $\Theta$ compares the cost of stretching the segment with the cost of bending it, which depends on how strongly the two are geometrically coupled. For small $\Theta$, stretching is energetically cheaper and the ground strain is transmitted as axial strain to the fiber. However, for large $\Theta$, bending is cheaper and the strain is absorbed.
Evaluating the integrals for the doubly-clamped mode shape of Equation~\ref{eq:mode_shape} gives $C_1 = \pi^2/2L$ and $C_2 = 2\pi^4/L^3$.
Inserting these, together with $A = \pi r^2$ and $I = \pi r^4/4$, all of the cable's material properties and the segment length cancel, and the parameter reduces to the compact, physically interpretable form
\begin{equation}
\label{eq:theta_sag}
\Theta = \frac{1}{2} \left( \frac{q_{f0}}{r} \right)^2.
\end{equation}
\textit{The static strain transfer efficiency thus depends entirely on the ratio of the segment's initial sag to its radius, irrespective of what produced that sag: small sag relative to the cable's thickness yields near-complete axial strain transfer, while large sag compared to the radius causes the segment to accommodate endpoint displacement through bending rather than through axial elongation.
The radius enters because it sets the ratio of axial to bending stiffness, which scales as $1/r^2$. As a practical threshold, keeping the loss in transfer efficiency below about 3\% requires $\Theta < 1/32$, corresponding to a sag smaller than about one quarter of the cable radius, $q_{f0}/r < 1/4$.}

\subsection{Origin of the Initial Sag}
\label{subsec:what_sets_sag}

Equation~\ref{eq:theta_sag} determines the strain transfer efficiency from the initial deflection, but leaves the origin of that deflection unconstrained.
This initial sag $q_{f0}$ can be produced either by gravitational sagging of an otherwise straight cable, or by curvature retained by the cable itself. Both are present in any real deployment. The estimates below establish how each mechanism scales with the segment length, and which of the two is expected to dominate.

\subsubsection{Gravity-induced sag}
Consider first a cable without retained curvature whose load-free shape is straight, so that the sag is produced only by its own weight. The total potential energy of the segment, measured relative to the straight configuration, is $\Pi = U_f + V_g$, where the bending energy is $U_f = \frac{1}{2} EI C_2 q_f^2$ and $V_g$ is the gravitational potential energy. With no initial axial tension the axial energy vanishes at rest. Gravity acts as a uniform distributed load $\rho A g$ along the segment, so
\begin{equation}
V_g = -\int_0^L \rho A g\, w(x)\, dx = -\frac{1}{2} \rho A g L\, q_f = -\frac{1}{2} F_g q_f,
\end{equation}
where $F_g = \rho A g L$ is the total gravitational force on the segment. Equilibrium occurs at the minimum of the potential energy, $\mathrm{d}\Pi/\mathrm{d}q_f = 0$, which gives
\begin{equation}
\label{eq:xf0}
EI C_2 q_{f0} = \frac{1}{2}F_g \quad \Rightarrow \quad q_{f0} = \frac{\rho A g L}{2EI C_2} = \frac{\rho A g L^4}{4 \pi^4 EI}.
\end{equation}
This analytical result of gravitationally induced sag is validated through comparison against the numerical model in Text S1 in Supporting Information S1.

Substituting $q_{f0}$ into Equation~\ref{eq:theta_sag} expresses $\Theta$ entirely through material and geometric quantities,
\begin{equation}
\label{eq:theta_grav}
\Theta_\mathrm{grav} = \frac{\rho^2 g^2 A^3 L^8}{128 \pi^8 E^2 I^3}.
\end{equation}
The extremely strong dependence on the segment length, $\Theta \propto L^8$, follows from the gravitational sag itself scaling as $L^4$.

\subsubsection{Sag from retained curvature}
Cables are stored and transported on spools and do not necessarily relax to a fully straight shape when unrolled. 
We estimate the resulting sag over a given span by treating the segment as an arc of uniform curvature radius $R$. This is a back-of-the-envelope description and not the real shape a cable takes, since a uniformly curved cable would close into a circle and its arc is not the mode shape assumed in Equation~\ref{eq:mode_shape}. 
Because a real cable retains curvature that varies along its length, we use this simple description to estimate how the sag scales with the span, not to predict the sag of a deployed cable.
The largest deviation from the straight line occurs at the midspan and is
\begin{equation}
\label{eq:sag_bend}
q_{f0} \approx \frac{L^2}{8R},
\end{equation}
for $L \ll R$, so that
\begin{equation}
\label{eq:theta_bend}
\Theta_\mathrm{bend} \approx \frac{L^4}{128\, R^2 r^2}.
\end{equation}
Unlike for the gravitational case, the midspan sag is determined only by the segment length and the retained curvature radius. The Young's modulus, the density and the gravitational acceleration do not influence the sag, although they may still act indirectly by influencing both the curvature a cable retains and the spacing of its contact points. The dependence on segment length is also much weaker, $\Theta \propto L^4$ instead of $L^8$. Inverting the coupling criterion $\Theta < 1/32$ gives the longest span such a segment can bridge while still transferring strain efficiently,
\begin{equation}
\label{eq:Lmax}
L_\mathrm{max} \approx \sqrt{2\,r\,R}.
\end{equation}
Note that $R$ is the curvature which the cable retains after unrolling. This is generally substantially larger than the radius of the spool it was stored on, and it is a result of the cable's construction and handling history and not a property obtainable from a datasheet.

\subsubsection{Which mechanism dominates}
Since the gravitational sag grows as $L^4$ and the retained-curvature sag only as $L^2$, retained curvature dominates at short spans and gravity must eventually take over at long ones. Setting Equation~\ref{eq:xf0} and Equation~\ref{eq:sag_bend} equal gives the span at which the two contribute equally,
\begin{equation}
\label{eq:Lstar}
L^{*} = \pi^2 \sqrt{\frac{EI}{2 R \rho A g}}.
\end{equation}
For representative cable parameters ($E = 4$~GPa, $r = 2$~mm, $\rho = 1200$~kg\,m$^{-3}$) and a retained curvature radius of $R = 2$~m, Equation~\ref{eq:Lstar} gives $L^{*} \approx 2.9$~m. Suspended segments in a surface deployment are much shorter than this, which leads to the expectation that the initial sag of a real draped cable is governed by retained curvature rather than by gravity, and consequently that $\Theta$ cannot be predicted from cable properties and gravity alone. This expectation is tested against measurements in Section~\ref{subsec:initial_sag}.

\section{Methods}
\subsection{Numerical Methods}
\label{subsec:numerical_methods}

The analytical treatment above relies on several approximations: small slopes, a reduction of the deformation to two modes, negligible shear, and a static configuration. We complement it with a numerical model, which both tests these approximations and extends the analysis to the frequency-dependent response.
To this end, we model the suspended cable segment as a 1D Timoshenko beam \cite<e.g.,>[]{weaver_vibration_1990}. 
This theoretical formulation assumes linear elasticity and incorporates finite rotations and shear, allowing for the accurate simulation of large geometric deformations and coupling between transverse bending and axial elongation.

The model is implemented in the open-source finite element framework FEniCS \cite{logg_automated_2012}, with the segment of Figure~\ref{fig:coupling_and_modes}b discretized into 40 beam elements. 
The load-free configuration of the beam is a circular arc of curvature radius $R$, representing the curvature the cable retains after unrolling, and gravity acts on it as a distributed load.
The initial shape of the segment is calculated first, using a nonlinear Newton solver. The equilibrium shape carries a sag produced by the retained curvature, by gravity, or by the two together, following the treatment in Section~\ref{subsec:what_sets_sag}. 
The static strain transfer efficiency is then determined by displacing the endpoints from this equilibrium and comparing the resulting change in the axial length of the beam with the imposed change of span. The applied displacement corresponds to a nominal strain of 1 $\,\mu\varepsilon$,
which is small enough that the behavior is in the linear regime that the analytical model treats.
Finally, the model is linearized about the equilibrium shape, and the dynamic behavior is characterized by solving for the natural vibration modes and by a steady-state frequency response analysis. Dissipation is modeled as a complex stiffness $K(1 + i\beta)$ with loss factor $\beta = 0.1$.

The parameters governing the behavior of the beam, and the ranges over which they were varied, are listed in Table~\ref{tab:parameters}. Each parameter was sampled at three geometrically spaced values. Gravitational acceleration and retained curvature were additionally sampled at the limiting values $g=0$ and $R=\infty$. All 1296 combinations were modeled. 

\begin{table}[htb]
\caption{Parameters of the beam model and the values over which they were varied. 
The shear modulus is not varied independently but follows from the Young's modulus and an assumed Poisson ratio $\nu = 0.35$ as $G = E/[2(1+\nu)]$, applied with a shear correction factor $\kappa = 6/7$. Internal shear is negligible for a beam as slender as a cable segment, so the results are insensitive to this choice.}        \label{tab:parameters}
\begin{tabular}{llll}
\hline
Parameter & Symbol & Unit & Values \\ \hline
Cable radius                & $r$    & mm             & 0.5, 1.12, 2.5 \\
Segment length              & $L$    & cm             & 5, 10, 20 \\
Young's modulus             & $E$    & GPa            & 1, 5.48, 30 \\
Density                     & $\rho$ & kg\,m$^{-3}$   & 700, 954, 1300 \\
Gravitational acceleration  & $g$    & m\,s$^{-2}$    & 0, 1, 3.16, 10 \\
Retained curvature radius   & $R$    & m              & 0.25, 1.12, 5, $\infty$ \\ \hline
\end{tabular}
\end{table}

A second set of simulations reproduces the laboratory configurations directly, so that the model and the measurements can be compared at the larger laboratory amplitudes. Each of the 56 configurations is modeled with the sag measured for that segment prescribed as its initial shape, and with the cross-section and the Young's modulus measured for that cable. Gravity is set to zero in these runs, because the initial shape is taken from the measurement rather than produced by the model, and adding gravity would deflect the segment further. Each configuration is run at the strain amplitudes which were imposed at the endpoints of that configuration in the experiment (reported in Figure S3 in Supporting Information S1), in extension and in compression for each case.

\subsection{Experimental Methods}
\label{subsec:experimental_methods}

The analytical model predicts the strain transfer efficiency of a single suspended segment from its initial sag and its radius alone. We test this prediction experimentally, by clamping individual cable segments of measured geometry, imposing a known change of the span between the clamps, and measuring the resulting axial strain of the segment. 
The setup realizes the assumptions of the model. Each segment is suspended over a free span between two clamps. Clamping enforces the no-slip condition, and it is the boundary condition underlying the mode shape of Equation~\ref{eq:mode_shape}. One of the two clamps is mounted on a shaker and driven along the axis of the segment, so that the strain is imposed at the endpoints and in line with the cable, as a ground strain is for a segment much shorter than the seismic wavelength. The initial sag is set by the way each segment is mounted and is measured directly, since the model treats it as a given property of the segment irrespective of its origin.
Because $\Theta$ depends on the sag and the radius, we vary the two quantities through the choice of cable and through the span and the mounting state. The resulting configurations cover 5 decades of $\Theta$.
The seven cables also span an order of magnitude in Young's modulus and more than a factor of two in density. Neither quantity enters $\Theta$, so the model predicts that neither affects the strain transfer directly. Covering a range of both makes that prediction testable rather than assumed. 

\begin{figure}[!tb]
    \centering
    \includegraphics[width=\textwidth]{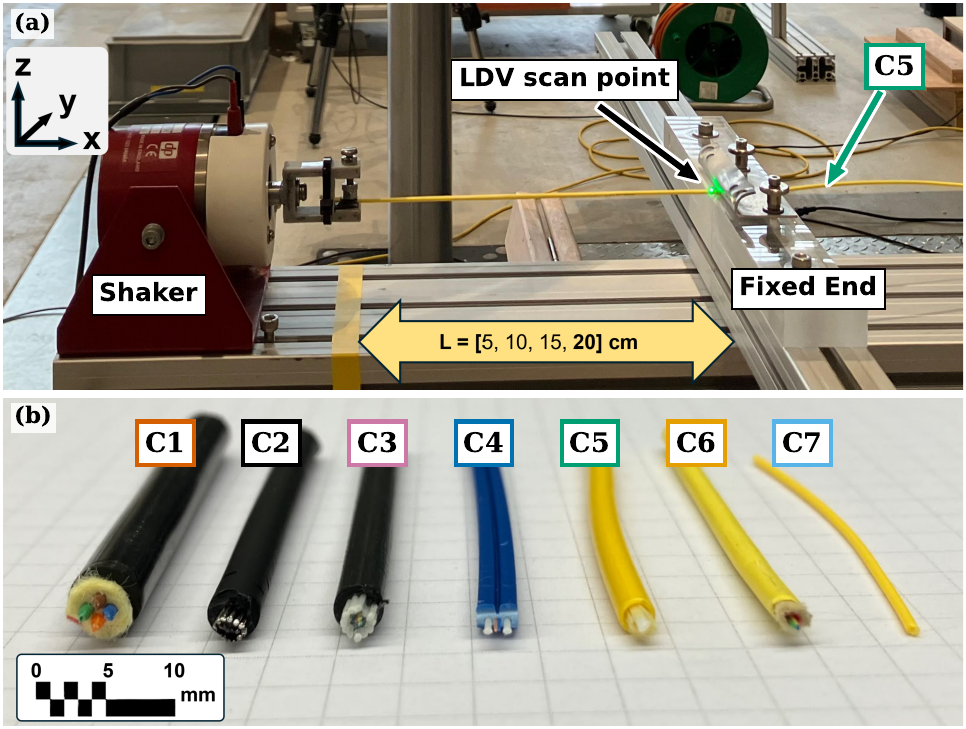}
    \caption{\textbf{Segments of seven fiber-optic cables are suspended between two clamps and strained by a shaker that drives one clamp along the cable axis.}
    \textbf{(a)} The setup, with the shaker and its clamp on the left, the free span in the center, and the fixed clamp on the right. The span $L$ is set by the position of the movable cross profile. The green dot on the cable is one of the scan points of the laser Doppler vibrometer, which measures the cable motion without contact from above the setup.
    \textbf{(b)} Cross-sections of the seven cables, ordered by outer diameter (Table~\ref{tab:cables}).}
    \label{fig:setup}
\end{figure}

The setup is shown in Figure~\ref{fig:setup}a. 
A Data Physics GW-V4/PA30E shaker is bolted to one end of an aluminum profile with its head leveled and driven along the profile axis. 
One end of the cable is clamped to the shaker with a 3D-printed connector, the other end is clamped to a cross profile whose position along the aluminum profile sets the span $L$ between the two clamps. 
Seven cables were measured at spans of L = 5, 10, 15 and 20~cm, each in two initial states. In the taut state the cable was pulled straight before the clamp was closed. In the second state, 1 to 2~mm of cable was pushed back through the clamp after straightening, giving the segment additional intended sag. This yields 56 configurations. 
The seven cables span a wide range of construction and outer diameter (Table~\ref{tab:cables}). They are labeled C1 to C7 in order of decreasing outer diameter. 
The Young's modulus was determined from three tensile tests per cable on a Zwick Roell Z010 testing machine.
The Young's modulus and the density are reported because they enter the sag that gravity alone would produce (Equation~\ref{eq:xf0}), against which the measured sags are compared in Section~\ref{subsec:initial_sag}, and because they parameterize the numerical model.
C4 is a flat ribbon instead of a round cable, and its dimensionless parameter is evaluated from its rectangular cross-section (Equation~\ref{eq:theta}), with the segment bending across its 1.9~mm thickness.

\begin{table}[!tb]
\caption{The seven fiber-optic cables used in the experiments, ordered by outer diameter. Outer dimensions are the diameter, except for the flat ribbon C4, for which the thickness and width are given. All parameter values were obtained through measurements. The Young's modulus is the mean of three tensile tests with its standard deviation.}
\label{tab:cables}
\small
\begin{tabular}{l p{3.3cm} p{4.3cm} l l l}
\hline
ID & Designation & Construction & Dim. [mm] & $\rho$ [kg\,m$^{-3}$] & $E$ [GPa] \\ \hline
C1 & Silixa PS-2S2M-1PU055-01 & Two multi-mode and two single-mode fibers, tight-buffered, aramid yarn, polyurethane jacket & 5.5 & 1050 & 8.9 $\pm$ 0.8 \\
C2 & Manufacturer unknown & Gel-filled steel wire braid in a hard plastic shell & 3.8 & 2270 & 28.6 $\pm$ 3.6 \\
C3 & Silixa BRUsens PS-48-2PE038-01-B & Glass wool, gel-filled, hard plastic shell & 3.8 & 1160 & 11.5 $\pm$ 1.2 \\
C4 & FIBRASENS 2OS2-FS02-19/31-T-BU & Two individually jacketed fibers joined as a flat ribbon, LSZH & 1.9 $\times$ 3.1 & 1530 & 4.2 $\pm$ 0.2 \\
C5 & Corning SMF-28 Ultra, 3\,mm OFNR & Tight-buffered core, loose aramid yarn, flame-retardant jacket & 2.9 & 960 & 2.8 $\pm$ 0.2 \\
C6 & HUBER+SUHNER G657.A2 LSFH & Four fibers in a tube, aramid yarn, LSZH jacket & 2.3 & 1190 & 6.8 $\pm$ 0.5 \\
C7 & Thorlabs SMF-28-100 & Single fiber in a thin Hytrel tube & 0.9 & 1240 & 2.1 $\pm$ 0.2 \\ \hline
\end{tabular}
\end{table}

The segments were measured with a three-component scanning laser Doppler vibrometer (LDV) (Polytec PSV-500 Scanning Vibrometer). Three beams at different incidence angles give the full particle-velocity vector at each scan point with a non-contact measurement. A scan therefore yields both the static three-dimensional position of every point along the segment, from which the initial geometry is obtained, and the three-component velocity time series at that point, from which the deformation is obtained.
Scan points were placed along a line on the upper surface of the cable, with the outermost point about 2~mm inside each clamp. The 10, 15 and 20~cm spans were sampled with 17 points and the 5~cm span with 9. The cables were sprayed with chalk to increase reflectivity, and C7 required a strip of reflective tape along its length because it was too thin to place scan points on it reliably. 
Data were recorded at 5~kHz over 12~s, with five repeated measurements median-stacked at each point. The acquisition was repeated for each scan point.
For every configuration, one additional scan point recorded the three-component velocity of the shaker head. 
A separate acquisition recorded the three-component velocity of the shaker head with no cable attached, to identify resonances of the shaker itself.
The accuracy of the measured geometry is limited mainly by the placement of the scan points. Because the scan line runs along the top of a round cable, a point placed slightly off the crest is displaced vertically as well as laterally. The sag is obtained from a fit of the mode shape to all scan points of a segment (Section~\ref{subsec:data_processing}), so such placement errors are largely averaged out rather than entering through a single point. 

The excitation signal was a logarithmic sweep from 1 to 500~Hz over 10~s. 
The resulting strains applied at the segment endpoints in the quasi-static frequency band (1-25 Hz) range from about 3 to 4000~$\mu\varepsilon$ across the 56 configurations. All configurations were driven with the same input signal. The imposed strain nonetheless differs between them, because the displacement of the shaker head depends on the response of the segment, so that axially compliant segments are strained more than stiff ones.
These amplitudes are up to three orders of magnitude larger than the peak strains of large regional and teleseismic earthquakes, which reach at most about 1 $\mu\varepsilon$ \cite{agnew_dynamic_2014}. Such large amplitudes were necessary to obtain a sufficient signal-to-noise ratio from the LDV velocities when converted to strain. Whether the response remains linear over this range, and the implications for the comparison with the model, are addressed in Section~\ref{subsec:data_processing}.

\subsection{Data Processing}
\label{subsec:data_processing}

The velocities are integrated in the frequency domain to obtain displacements.
The division is capped below 0.5~Hz so that the $1/f$ factor does not amplify low-frequency noise. All quantities below follow from these displacements and from the static positions.
The static scan-point positions give the initial shape of each segment. 
The two outermost points define the chord and each point is described by its coordinate along the chord and its offset perpendicular to it.
The dominant direction of these perpendicular offsets is obtained as their first singular vector, and the shape of the segment is the profile obtained by projecting the offsets onto the dominant direction.
The measured sag, $q_{f0}$, is the amplitude of the mode shape of Equation~\ref{eq:mode_shape} in this profile, obtained by a one-parameter least-squares fit. 
The dimensionless parameter follows from the measured sag and the cable cross-section as $\Theta = A\,q_{f0}^2/(8I)$ (Equation~\ref{eq:theta}), which reduces to $\frac{1}{2}(q_{f0}/r)^2$ for cables with a circular cross-section.

The strain transfer efficiency compares the elongation of the cable with the elongation imposed on its endpoints. 
The reference elongation $\delta L$ is the change in the straight-line distance between the two endpoints. The cable elongation $\delta q_l$ is the change in arc length of the segment, which is obtained by summing, over the sections between adjacent scan points, the component of the relative displacement of their ends along the local section direction. 
The ratio $\delta q_l / \delta L$ is the fraction of the imposed axial strain that is transferred to the segment.
The strain transfer is evaluated in the frequency domain over the full sweep record, with $X = \delta L$ as input and $Y = \delta q_l$ as output. Auto- and cross-spectral densities $S_{XX}$, $S_{YY}$, and $S_{XY}$ are estimated by Welch's method, using 2~s Hann-tapered segments with 50\% overlap, which gives $\approx$ 11 averages at a resolution of 0.5~Hz. The transfer function and the coherence are
\begin{equation}
\label{eq:H1coh}
H_1(f) \;=\; \frac{S_{XY}(f)}{S_{XX}(f)},
\qquad
\gamma^2(f) \;=\; \frac{\left|S_{XY}(f)\right|^2}{S_{XX}(f)\,S_{YY}(f)} .
\end{equation}
The $H_1$ estimator is the least-squares solution for noise on the output. 
The coherence measures the fraction of the output power at a given frequency that is linearly explained by the input. 
The quasi-static strain transfer efficiency $\eta$ of a configuration is computed as the mean of $|H_1|$ over the band from 1 to 25~Hz, using only the frequency bins in which the coherence exceeds 0.7, to remove bins where the estimate is not reliable.
Two quantities are reported alongside $\eta$, both evaluated over the quasi-static band:
the standard deviation of $|H_1|$, and the mean coherence. Together they indicate how robust the evaluated $\eta$ is for a given configuration. A large standard deviation means that the transfer function is not flat across the band, either because the response there is not quasi-static or because the estimate is noisy, and a low coherence indicates noisy data.

The analytical efficiency requires the change in sag during a cycle to remain small compared with the initial sag, which the large laboratory strain amplitudes (Section~\ref{subsec:experimental_methods}) do not guarantee. We therefore assess the effect of the large experimental amplitudes.
As described in Section~\ref{subsec:numerical_methods} we model the laboratory configurations at their measured drive amplitudes. The results are processed in a comparable manner as the measurements. The displacements are projected onto the direction of the static segment, as for the measured $\delta q_l$, which gives the efficiencies $\eta_\mathrm{ext}$ and $\eta_\mathrm{comp}$, for extensional and compressive strains respectively. Their mean is compared with the small-amplitude prediction. 
Additionally, we assess whether the predicted asymmetry is present in the measurements. The standard efficiency estimate at the drive frequency removes the asymmetry, so the efficiencies are instead estimated in the time domain.
The samples are split into extension ($\delta L > 0$) and compression ($\delta L < 0$), and the efficiencies are determined for both groups. Details and results of this analysis are given in Text S3.3 in Supporting Information S1.

The dynamic response of each segment is described by two transfer functions relative to the same input $X = \delta L$: the strain transfer function $H_\eta = H_1$ of Equation~\ref{eq:H1coh} with output $Y = \delta q_l$, and a midspan transfer function that measures how much the midspan moves and thus the amplitude of the first bending mode of the segment. 
Because the static bend of the segment is not necessarily vertical (Section~\ref{subsec:initial_sag}), the direction in which the midspan moves is not known in advance. The displacement of the scan point closest to midspan is therefore projected onto two orthogonal directions perpendicular to the chord and an $H_1$ estimate $H_{\perp,i}$ with coherence $\gamma^2_{\perp,i}$ is formed for each direction ($i = 1, 2$). The two are combined into the total transverse midspan motion and its coherence,
\begin{equation}
\label{eq:Hmid}
\left|H_\mathrm{mid}(f)\right| = \sqrt{\left|H_{\perp,1}(f)\right|^2 + \left|H_{\perp,2}(f)\right|^2},
\qquad
\gamma^2_\mathrm{mid}(f) = \frac{\left|S_{X\perp,1}\right|^2 + \left|S_{X\perp,2}\right|^2}{S_{XX}\left(S_{\perp\perp,1} + S_{\perp\perp,2}\right)} ,
\end{equation}
both of which are independent of the direction of the bend. The coherence $\gamma^2_\mathrm{mid}$ is the fraction of the transverse midspan motion that is linearly explained by $\delta L$.
Resonances are identified as peaks of $|H_\mathrm{mid}|$.
The amplitude is normalized to its maximum over the bins with $\gamma^2_\mathrm{mid} \ge 0.7$, and peaks are retained if they are separated by at least 10~Hz, and their coherence $\gamma^2_\mathrm{mid}$ is at least 0.7. 
The transfer functions do not separate the response to axial endpoint motion from the response to transverse endpoint motion. Where the shaker head moves transversely, the segment follows this motion while $\delta L$ remains small, and $|H_\mathrm{mid}|$ shows a peak without a resonance of the segment. To identify such peaks, the direction of the input is quantified from the shaker-head recording of each configuration by the ratio of transverse to axial velocity amplitude $R_{sh}$.     
Peaks with $R_\mathrm{sh} > 1$, where the shaker head moves more transversely than axially, are classified as driven by transverse shaker motion.

\section{Results}

\subsection{Initial Cable Sag}
\label{subsec:initial_sag}

\begin{figure}[!tb]
    \centering
    \includegraphics[width=\textwidth]{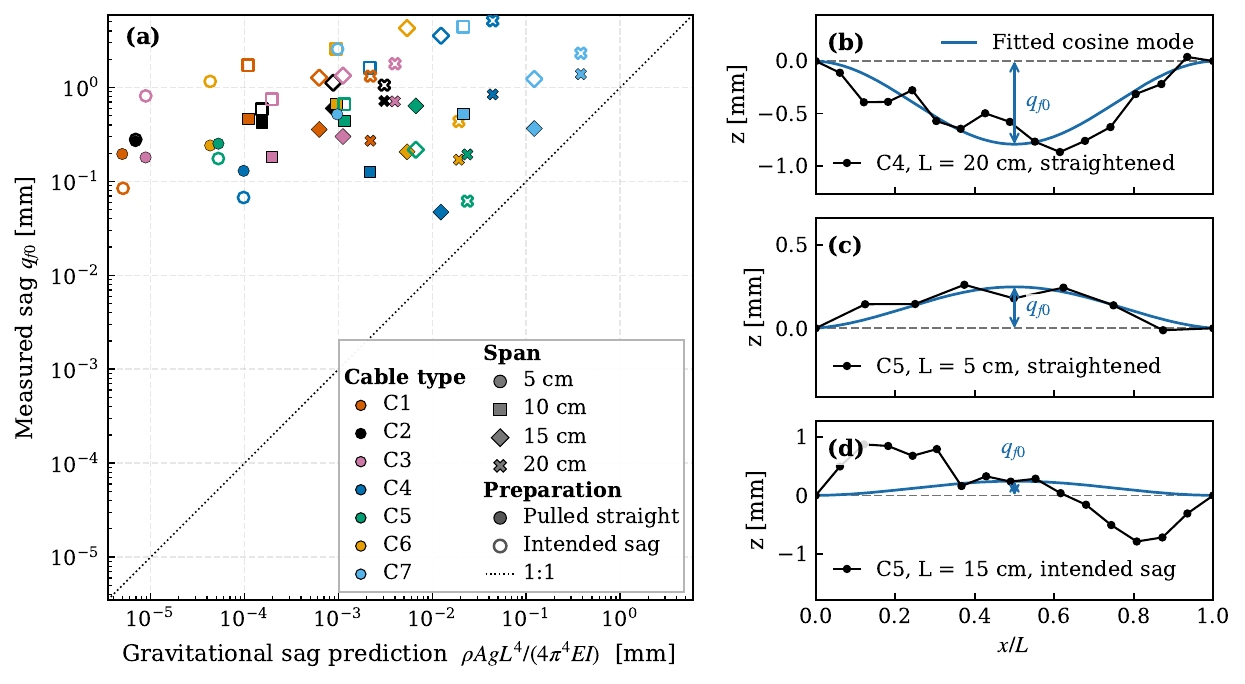}
    \caption{\textbf{The sag of a suspended segment is set by retained curvature rather than by gravity, exceeding the gravitational prediction by orders of magnitude and deviating from the downwards, single-lobed shape gravity would produce.}
    \textbf{(a)} Measured midspan sag $q_{f0}$ of the 56 configurations against the sag predicted for gravity acting on an initially straight segment (Equation~\ref{eq:xf0}). Color indicates cable type and marker shape indicates span. Filled symbols are the configurations pulled straight before the measurement, open symbols those given intentional sag. The dotted black line marks equality of measurement and prediction. All configurations lie far above the 1:1 line.
    \textbf{(b-d)} Measured segment profiles of three configurations in side view with fitted mode:
    \textbf{(b)} a single-lobed bend, \textbf{(c)} a single-lobed bend with a bending direction upwards instead of downwards, and \textbf{(d)} a segment that is not single-lobed.}
    \label{fig:initial_sag}
\end{figure}

The midspan sag, $q_{f0}$, was determined for all 56 configurations and ranges from 0.1 to 4~mm. Figure~\ref{fig:initial_sag}a compares the measured sags against the sags the same segments would take under gravity if their load-free shapes were straight (Equation~\ref{eq:xf0}).
Half of the configurations were intentionally pulled straight before the measurement, and their sags still exceed the gravity prediction by one to four orders of magnitude. 
In the other half, sag was introduced deliberately to cover a wider range of $\Theta$, and these consequently have even larger sags.
The measured profiles scatter little about the fitted mode shape compared with the sag (Figure~\ref{fig:initial_sag}b-c), and the intended-sag configurations consistently show larger sags than the taut ones. Both indicate that the measured sags reflect the geometry of the segments rather than errors in scan-point placement.
Because the sag of each segment was set by how it was mounted and not by free draping, these measurements do not test the predicted dependence of either mechanism on the segment length.

The orientation and shape of the measured bends are also inconsistent with a gravitational origin. The bends are not purely vertically downwards, and not all profiles are single-lobed (Figure~\ref{fig:initial_sag}b-d).
While gravity acting on a straight segment would produce a symmetric shape in the vertical plane, a retained bend has no set orientation and no prescribed shape.
The initial shape of a suspended segment is therefore governed by the curvature the cable retains from spooling and handling, with gravity as a much smaller secondary contribution.
The sag cannot be predicted from cable properties and gravity, and $\Theta$ is evaluated from the measured sag in the following.

\subsection{Static Strain Transfer Efficiency}
\label{subsec:static_ste}

\begin{figure}[!tb]
    \centering
    \includegraphics[width=\textwidth]{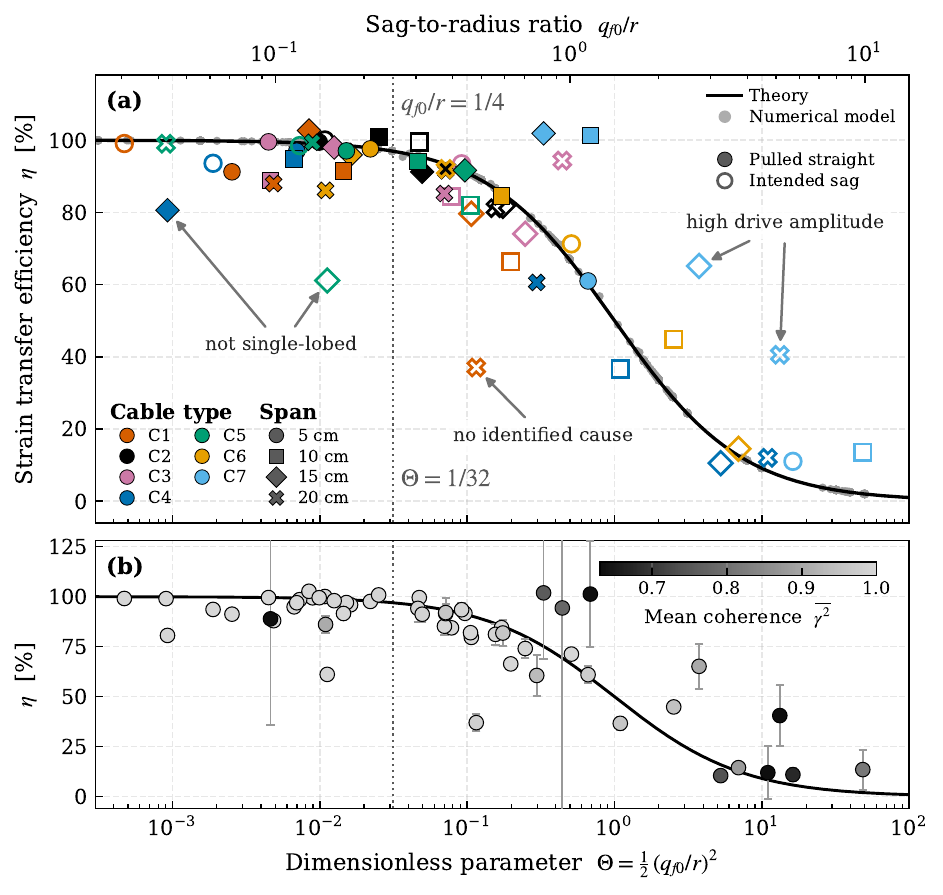}
    \caption{\textbf{The static strain transfer efficiency of a suspended segment is governed by the ratio of its sag to its radius, with the analytical prediction, the numerical simulations and the laboratory measurements in agreement.} 
    \textbf{(a)} Strain transfer efficiency $\eta$ against $\Theta$. The solid line is the analytical prediction of Equation~\ref{eq:strain_transfer}. 
    Small grey markers are numerical modeling results which lie on the analytical curve and are largely hidden by it. 
    Colored markers are the laboratory configurations, with color indicating cable type, marker symbol indicating span, and filled and open symbols distinguishing the taut and intended-sag configurations. The upper axis gives the corresponding sag-to-radius ratio, and the dotted line marks $\Theta = 1/32$, equivalently $q_{f0}/r = 1/4$, below which the predicted efficiency loss stays within 3\%. 
    The outlier configurations discussed individually in the text are pointed out.
    \textbf{(b)} Quality assessment of the measurements. The same measurements and the same axes, with marker shading giving the mean coherence of the transfer function over the quasi-static band and vertical bars the standard deviation of the efficiency across that band. Low coherence and large standard deviation both indicate that the measured efficiency is not stable across the quasi-static band.}
    \label{fig:static_strain_transfer}
\end{figure}

The analytical model predicts that the static strain transfer efficiency of a suspended segment depends only on the ratio of its initial sag to its radius, through $\eta = 1/(1+\Theta)$ with $\Theta = \frac{1}{2}(q_{f0}/r)^2$ (Equation~\ref{eq:theta_sag}).
The numerical results match the prediction (Figure~\ref{fig:static_strain_transfer}). 
Plotted against $\Theta$ evaluated from the resulting sag, all of the modeled configurations collapse onto the analytical curve.
The measurements scatter around the same curve. For each of the 56 configurations, $\Theta$ was evaluated from the measured midspan sag and the cable radius, and the strain transfer efficiency was obtained from the coherence-gated transfer function in the quasi-static band (Section~\ref{subsec:data_processing}). For one configuration (C7, taut, spanning 20~cm) none of the frequency bins reached the minimum coherence of 0.7, and it is therefore absent from Figure~\ref{fig:static_strain_transfer}.

The measurements follow the curve more closely at low $\Theta$ than at high $\Theta$, which follows from the shape of the curve. In the flat region below the threshold, an error in the measured sag moves a configuration along the $\Theta$-axis without changing the predicted efficiency a lot. Where the curve is steep, the same relative error in the sag produces a larger change in the predicted efficiency. Below the coupling threshold $\Theta < 1/32$, for which the model predicts an efficiency above 97\%, all configurations transferred at least 87\% of the applied strain excluding the two configurations with double-lobed mode shape discussed as outliers below. 

\subsubsection{Response at the laboratory strain amplitudes}
\label{subsec:linearity_results}

The analytical prediction and the numerical results characterize the response to small endpoint displacements. 
Seismic strains in the field fall in this regime, but the laboratory strains of up to 4000~$\mu\varepsilon$ do not necessarily, so the measured efficiencies cannot be assumed to equal the small-amplitude efficiency.

The model itself indicates how a large amplitude changes the response. Pulling the endpoints apart straightens the segment, which reduces its sag and raises its efficiency while pushing them together deepens the sag and lowers it. Once the change in sag during a cycle is no longer small compared with the initial sag, the segment therefore transfers strain more efficiently in extension than in compression. 
The deviations are opposite, such that their mean still recovers the small-amplitude efficiency for moderate asymmetry between compression and extension. 

The simulations of the laboratory configurations at their measured drive amplitudes show that the deviation is small for nearly all configurations. Their mean efficiency $\bar{\eta}$ lies within a few percent of the small-amplitude prediction. 
The exceptions are the two C7 configurations with intended sag at spans of 15 and 20~cm and the configuration excluded due to insufficient coherence, for which the simulated $\bar{\eta}$ exceeds the prediction by about 40\%.
They are marked in Figure~\ref{fig:static_strain_transfer}a and discussed with the outliers below.

The time-domain estimates from the measurements show the same trend as the numerical model, with higher efficiency in extension than in compression. They do not, however, match the simulated asymmetry configuration by configuration. Because this estimate is noisier than the transfer function, the comparison supports the direction of the effect but does not test its magnitude (Text S3 in Supporting Information S1).

\subsubsection{Outliers}
\label{subsubsec:outliers}

A number of configurations depart visibly from the curve.
The coherence and the standard deviation of the transfer function across the quasi-static band (Figure~\ref{fig:static_strain_transfer}b), together with the measured geometry and the drive amplitude, separate the outliers into three explicable groups and leave one outlier configuration unexplained (marked in Figure~\ref{fig:static_strain_transfer}a).

For segments whose initial geometry is not single-lobed but double-lobed, the correspondence between the midspan sag and the strain transfer is broken (Figure~\ref{fig:initial_sag}d).
The deviations from the chord have opposite signs and largely cancel, resulting in a small fitted sag. Consequently, $\Theta$ is small, which places the configuration to the left in Figure~\ref{fig:static_strain_transfer}a, where the model predicts near-complete transfer, while the segment still has a bending geometry to exploit, which is not captured by the single-lobe analytical model.
C5 at a 15\,cm span with intended sag (Figure~\ref{fig:initial_sag}d) and C4 at a 15\,cm span without intended sag are such cases.
Fitting their profiles instead with a double-lobed sine, $\phi(x) = \sin(2\pi x/L)$, and evaluating Equation~\ref{eq:theta} for this mode shape brings the prediction closer to the measurement: 71\% predicted against 61\% measured for C5, and 88\% against 81\% for C4.

A second group consists of configurations of dominantly C7, that exhibit low coherence and large standard deviations across the quasi-static band (Figure~\ref{fig:static_strain_transfer}b) indicating poor measurement quality. C7 is the thinnest cable of the set and required a reflective tape along its length for the LDV to be able to measure on its surface.
The tape is a candidate explanation for the reduced coherence, but we did not measure its effect. 

As a third group, the two C7 configurations with intended sag at spans of 15 and 20~cm lie above the curve due to their low axial stiffness which causes high drive amplitudes.
At these drive amplitudes the numerical model predicts a change in sag during a cycle large enough that the transfer function no longer recovers the small-amplitude efficiency. The measured efficiencies exceed the small-amplitude prediction by about 30 to 50\%, consistent with the numerical model evaluated at the same drive amplitudes (Text S3 in Supporting Information S1).

Finally, C1 at a 20~cm span with intended sag departs strongly from the curve without a low coherence, without a large standard deviation, and without an identifiable feature of its geometry. We have no explanation for the deviation of this configuration.

\subsection{Dynamic Strain Transfer Function}
\label{subsec:dynamic_ste}

At finite frequencies, the strain transfer is no longer a real fraction but a complex transfer function, with an amplitude that can exceed the static limit and a phase relative to the driving ground motion. 
We first describe the resonance behavior observed in the numerical simulations, and then comment on the experimentally measured frequency response functions.

\subsubsection{Predicted Dynamic Response}

The dynamic response of a suspended cable segment depends on its natural frequencies, most importantly the first flexural bending mode. This fundamental resonance frequency acts as a threshold dividing the cable's behavior into quasi-static, resonant, and inertia-dominated regimes. For a straight, horizontal Euler-Bernoulli beam with doubly clamped boundary conditions, the fundamental angular frequency is \cite<e.g.,>[]{weaver_vibration_1990}
\begin{equation}
\label{eq:omega1}
\omega_1 = \frac{22.4}{L^2} \sqrt{\frac{EI}{\rho A}}, \qquad f_1 = \frac{\omega_1}{2\pi}.
\end{equation}
If the seismic frequency band of interest lies sufficiently far below this resonance, the cable segment behaves in a predictable, frequency-independent manner.
While the fundamental frequency, $\omega_1$, is set by the beam properties, bending causes a geometric stiffening effect that increases the resonance frequency: a pre-bent cable exhibits a higher effective flexural stiffness than a perfectly straight counterpart. 

\begin{figure}[!tb]
    \centering
    \includegraphics[width=\textwidth]{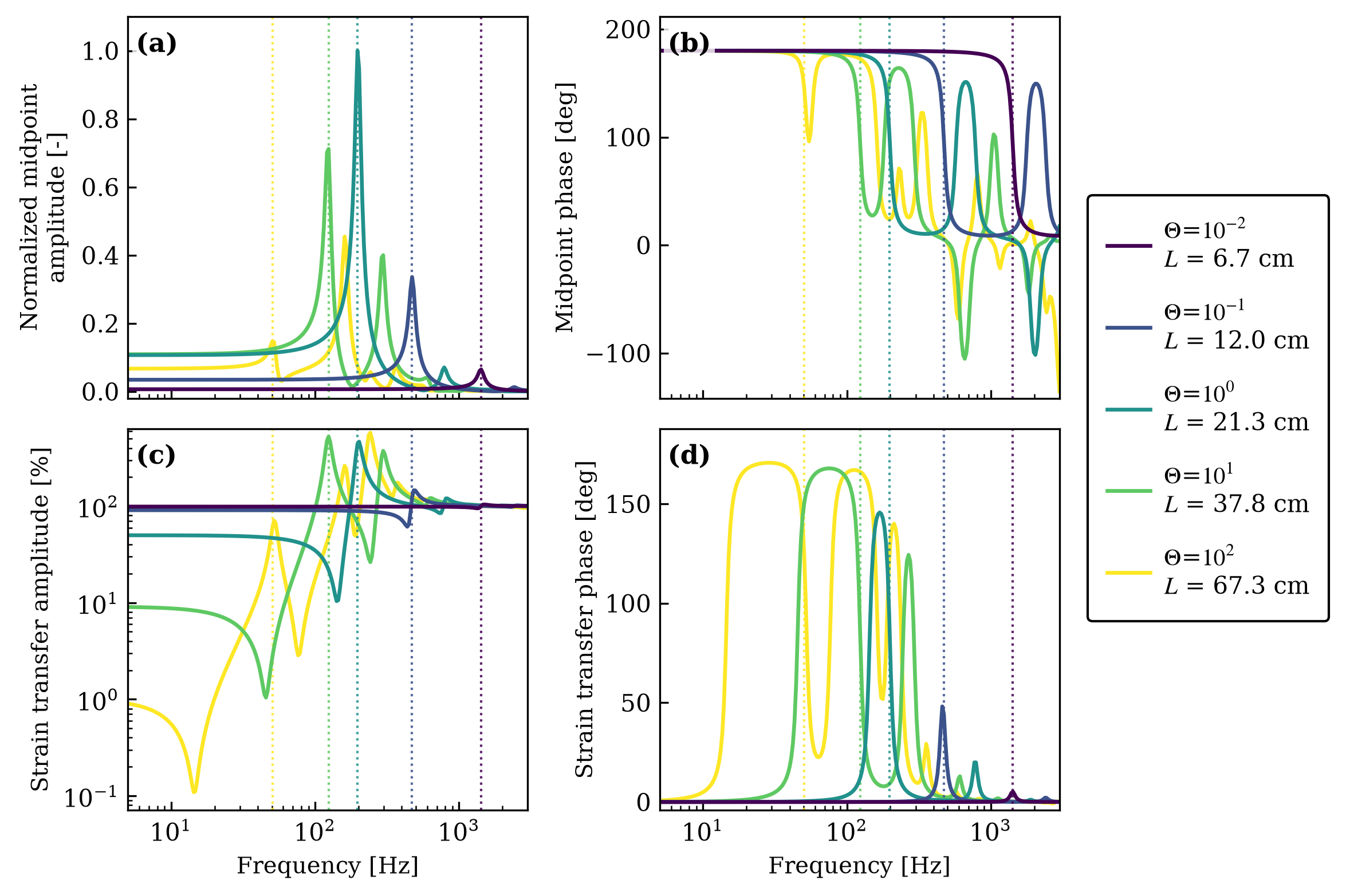}
    \caption{\textbf{Dynamic behavior of the axial and bending mode coupling and the resulting strain transfer.} Frequency response of a single suspended cable segment as the segment initial sag is varied to span a range of $\Theta$. Cable properties and retained curvature radius are fixed ($E = 4$~GPa, $\rho = 1200$~kg\,m$^{-3}$, $r = 2$~mm, $R = 2$~m) while for simplicity additional sagging due to gravity as a secondary effect is neglected ($g=0)$.
    The corresponding segment lengths are given in the legend. \textbf{(a, b)} Amplitude and phase of the cable's midspan deflection relative to its sagged state (not absolute deflection), measuring the intensity of the flexural mode during axial deformation. The deflection shows a typical resonance peak and a frequency-dependent phase shift. \textbf{(c, d)} Amplitude and phase of the dynamic strain transfer as a function of frequency. At low frequencies, the strain transfer matches the static limit. As the frequency approaches resonance, it is initially attenuated due to destructive mode coupling, followed by an amplification peak driven by constructive coupling as the phase rotates across the resonance. Above the resonance frequency, cable inertia suppresses the first flexural mode, resulting strain transfer approaching 100\%. Higher order modes produce additional peaks at higher frequencies. 
    Since larger $\Theta$ values here correspond to longer segments, they have lower resonance frequency (Equation~\ref{eq:omega1}). Deployments in the recommended range ($\Theta < 1/32$) have higher resonance frequencies and a response that is flat across the seismic band, as shown by the smaller $\Theta$ configurations plotted here.}
    \label{fig:frequency_response}
\end{figure}

To examine the dynamic response across the range of $\Theta$, we consider a cable segment of fixed material properties and cross-section and vary the suspended segment length, $L$, setting the initial sag of each configuration from a constant retained radius of curvature, $R$, through Equation~\ref{eq:sag_bend}. Increasing $L$ raises $\Theta$ and at the same time lowers the fundamental resonance frequency (Equation~\ref{eq:omega1}), so that the resonance peaks shift towards lower frequencies as $\Theta$ increases (Figure~\ref{fig:frequency_response}). The geometric stiffening
associated with the larger initial bend acts in the opposite sense, but does not
reverse this trend.

    At seismic frequencies well below the fundamental resonance, the cable responds quasi-statically, yielding a dynamic strain transfer that matches the static limit (Figure~\ref{fig:frequency_response}c). The driving ground motion is slow compared to the cable's mechanical response time, internal inertial forces remain negligible, and the segment transitions between equilibrium states in phase with the ground.
    How strongly the segment deviates from its quasi-static limit as frequency increases depends on $\Theta$. Below $\Theta \approx 1/32$, the strain transfer remains close to 100\% across the entire frequency band, and the amplitude and phase responses stay essentially flat (Figure~\ref{fig:frequency_response}c and d). For larger $\Theta$, the response becomes increasingly frequency-dependent as described in the following.
    
    As the driving frequency approaches the fundamental resonance, the ground strain strongly excites the bending mode, resulting in a peak of the midspan displacement amplitude at the resonance frequency. Just below the resonance, the midspan deflection is still close to in-phase with the ground movement. Consequently, ground displacement is absorbed into bending more effectively. This destructive coupling causes the strain transfer to drop below its static limit and reach a minimum.
    Traversing the resonance frequency, the phase of the midspan displacement undergoes an almost 180-degree phase shift, inverting how the bending mode and the axial mode interact. The cable starts moving in opposition to the ground driving motion: when the endpoints are pulled outward, the inertia of the cable causes it to bend further, rather than to straighten. The two effects add constructively: the fiber is stretched both by the ground and by its own whipping motion. This amplifies the signal, causing the strain transfer to spike immediately above the resonance frequency (Figure~\ref{fig:frequency_response}c).
    
    At driving frequencies significantly exceeding the first flexural mode, structural inertia ultimately suppresses the first flexural mode. The transverse flexure becomes ``too slow'' to follow the high-frequency oscillations, causing the midspan displacement to decay toward zero and the dynamic strain transfer to approach 100\%.
    However, higher order flexural modes of the segment produce additional resonance peaks and phase shifts at higher frequencies.

\subsubsection{Measured Dynamic Response}
\label{subsec:measured_dynamic}

Two properties of the experimental setup limit how quantitatively the measured dynamic response can be compared with the model results. First, the shaker head does not move purely along its drive axis: at certain frequencies where the shaker itself resonates it also moves transversely, introducing midspan displacements of the cable that are not the cable's own resonances.
Second, the shaker-side clamp is not fully rigid. The shaker head can be moved by the push-back of the cable which violates the clamped-clamped assumption.
The measured resonances are therefore properties of the coupled cable–shaker system and not those of an ideally clamped segment. The configurations identified as outliers in Section~\ref{subsubsec:outliers} are excluded from the following analysis.

\begin{figure}[!tb]
    \centering
    \includegraphics[width=\textwidth]{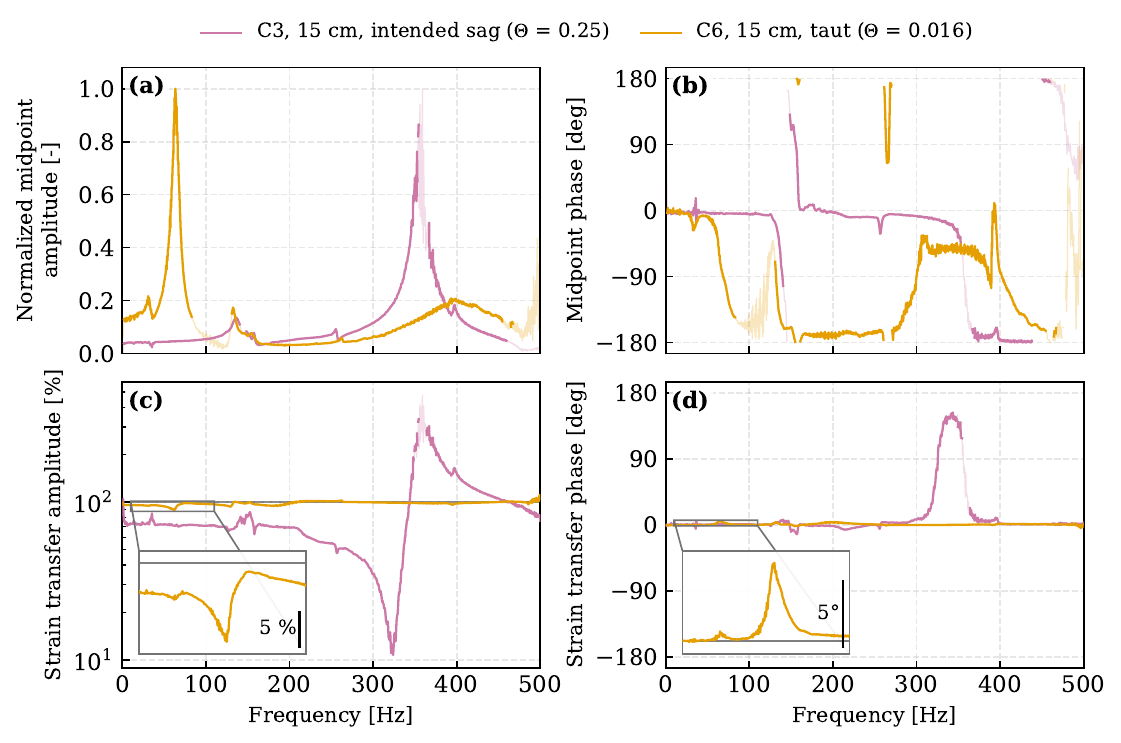}
    \caption{\textbf{The measured transfer functions show the dip and overshoot of the strain transfer around peaks of the midspan amplitude predicted by the model.}
    \textbf{(a, b)} Amplitude and phase of the midspan transfer function $H_\mathrm{mid}$, normalized to its maximum. \textbf{(c, d)} Amplitude and phase of the strain transfer function $H_\eta$. Configurations: C3, 15~cm, intended sag and C6, 15~cm, taut. Bins with coherence below 0.7 are shown with increased transparency.}
    \label{fig:measured_frf}
\end{figure}

The observed resonances in the measured transfer functions reproduce the qualitative structure predicted by the numerical model. Two examples are shown in Figure~\ref{fig:measured_frf}. At low frequencies the strain transfer is flat and in phase with the input. Approaching a peak of the midspan amplitude, the strain transfer drops below its quasi-static value, and immediately above the peak it rises above it, accompanied by a shift in the phase of the midspan and of the strain transfer. 

A large fraction of all cable segments exhibits a peak that is driven by transverse motion of the shaker head and not by the segment (all found resonances are shown in Figure S6 in Supporting Information S1). 
Most of these cluster between 115 and 150~Hz, independently of cable, span and initial state. This band contains the resonance of the bare shaker at 137~Hz. These peaks reflect the response of the cable to transverse motion of its endpoint and are not bending modes of the segment (Text S4.2 in Supporting Information S1).

The lowest remaining peaks behave like the first bending mode of the segment, but at lower frequencies than predicted. Their frequency decreases with increasing span, and for C1 at all spans and for C6 at the 10 and 15~cm spans the intended-sag configurations resonate at higher frequencies than the taut ones, consistent with the geometric stiffening of a bent segment. However, their frequencies lie well below the predictions of Equation~\ref{eq:omega1}. For C1 at spans of 10, 15 and 20~cm, the first modes are at 176, 89 and 54~Hz, a factor of 7 to 9 below the clamped prediction. For C5 at the same spans they are at 109, 74 and 48~Hz, a factor of 2 to 4 below the prediction (Text S4.3 in Supporting Information S1). 
We have no conclusive explanation for this deviation, but two contributions are possible. The first is the compliance of the shaker-side clamp. The endpoint of the cable at the shaker moves transversely at the resonances, so this end is not rigidly clamped, and a softer support lowers the resonance frequency. It may, however, not be sufficient on its own to explain deviations as large as those of C1, whose frequencies lie well below even the pinned prediction.
The second is the bending stiffness used in the prediction. The span, radius and density are measured with little uncertainty, whereas matching the measured frequencies of C1 would require a modulus about two orders of magnitude below the tensile value. The tensile modulus characterizes the axial stiffness of the cable, which is dominated by its strength members. For a layered cable whose components might slide relative to each other, the bending stiffness can be lower than the product of this modulus and the moment of inertia of the full cross-section.
The bending stiffness of the cables was not measured independently, so the two contributions cannot be separated with the present data.
Whatever their cause, the measured first modes are lower than predicted, so they provide a conservative estimate of where the dynamic response of a segment departs from the quasi-static one. For spans of up to 10~cm, all first modes lie above about 110~Hz. At a span of 20~cm the first modes lie between about 20 and 55~Hz, within the frequency band of many seismic applications.

\section{Discussion}
\label{sec:discussion}

\subsection{Sag-to-radius ratio governs strain transfer, and retained cable bend sets the sag}  

The strain transfer of a suspended segment is set by a single geometric ratio, the sag compared with the cable radius. Stiffness, density, segment length and gravity enter only through their effect on that sag, and the loss increases gradually once the sag exceeds roughly a quarter of the radius. 
The sag itself is not gravitational: measured values exceed the prediction for gravity acting on a straight cable by up to several orders of magnitude, even where the segment was pulled taut before testing. $\Theta$ is therefore a property of the cable's handling history and of the surface it is deployed on, not of its material specification, and it cannot be computed from a datasheet. 

Two main factors therefore set $\Theta$ in the field: the first is the curvature the cable retains after unrolling, which depends on the cable's material and construction, on the diameter of the spool and its storage history.
The second is the spacing of the ground contact points, which is set by the roughness of the surface together with how closely the cable conforms to it. A more compliant cable follows small undulations and touches down more often than a stiff one. $\Theta$ depends strongly on this spacing, since for a cable of given retained curvature the sag over a span grows as the square of the span (Equation~\ref{eq:sag_bend}). This exponent represents a back-of-the-envelope rather than a precise prediction, as a real cable does not retain a uniform curvature along its length. The two quantities are also not independent: a cable that retains curvature does not lie flat even on a smooth surface, so its own bend contributes to determining the contact point spacing. 

Bending stress relief offers a possible explanation for empirical observations that have so far lacked a mechanical interpretation. Because a draped cable sags between contact points and absorbs part of the ground strain as bending, our model predicts that unburied cables record lower amplitudes than buried ones, and that this loss is reduced for cables with larger radii. These predictions are consistent with the reduced amplitudes reported for unburied deployments \cite{an_traffic_2023, harmon_surface_2022, zandanel_earthquake_2026}, and with the observation that thicker cables exhibit better strain transfer \cite{probst_controlled_2026}, although other factors may also contribute. A direct quantitative comparison is not possible, however, since evaluating $\Theta$ requires the characteristic segment sag, which none of these studies measured or reported. 

For $\Theta$ to describe the strain transfer of a field measurement, the segment has to respond quasi-statically, which requires the seismic frequencies of interest to lie well below the segment's fundamental resonance. Near and above that resonance, the recorded strain is distorted in both amplitude and phase. The measured fundamental frequencies lie below the predictions for an ideally clamped beam, so Equation~\ref{eq:omega1} should not be relied on to place a deployment in the quasi-static regime. In our measurements, segments of up to 10~cm resonated above about 110~Hz, whereas segments of 20~cm resonated at frequencies between about 20 and 55~Hz. Because the fundamental frequency decreases and $\Theta$ increases with segment length, both requirements are satisfied by the same measure of keeping the segments short.
For the DAS measurement to be interpretable, this has to hold for every segment within a gauge length. A DAS channel averages the response of all segments it spans, so a single segment resonating inside the seismic band distorts the amplitude and phase of the whole channel while the remaining segments still respond quasi-statically. 

\subsection{Bending stress relief sets an upper bound on strain transfer} 

Bending stress relief, the mechanism we have identified and quantified here, may not be the only one limiting coupling in unburied deployments. Other mechanisms can act alongside it, including the frictional slip at the contact points introduced earlier. A back-of-the-envelope comparison of the friction available at the contact points with the force required to deform the cable suggests that slip is unlikely at the strain amplitudes of natural earthquakes, although the margin narrows for stronger active sources \cite{probst_controlled_2026}. Where such mechanisms are active, they would further reduce the transferred strain below the values predicted here, so our model gives an upper bound on the achievable strain transfer.

Our analysis also addresses only the first of the two transfer steps. The efficiency derived here concerns the ground-to-cable transfer, and we have assumed that the strain is then transmitted perfectly through the cable to the fiber. In practice this cable-to-fiber transfer can be imperfect, attenuating the strain through shear-lag mechanics. The strain reaching the fiber is therefore at most that predicted by $\Theta$, and may be lower depending on the cable.

\subsection{Limitations of the model}  
\label{subsec:limitations}

The derivation adopts a mode shape that satisfies clamped conditions at the contact points, prescribing the deflection as well as the slope. Clamping is the appropriate condition for a segment bent in the vertical plane, where the cable cannot pass through the surface and must be tangent to it at the contact points. A bend in the horizontal plane is not constrained in this way, since friction fixes the position of the contact point but not the tangent, and pinned conditions are then the better description. A cable that retains curvature carries its bend at an arbitrary orientation, so real segments are expected to behave somewhere between the two cases. 
The pinned case follows from the same derivation by replacing the mode shape with a half period, $\phi(x) = \sin(\pi x / L)$, and re-evaluating the two integrals. This leaves $C_1 = \pi^2/2L$ unchanged and reduces $C_2$ from $2\pi^4/L^3$ to $\pi^4/2L^3$, so that $\Theta = 2\left(q_{f0}/r\right)^2$, four times its clamped value at the same sag. The coupling criterion becomes more strict with a maximum allowable sag of $1/8$ of the cable's radius. 
The admissible sag of a real segment therefore lies between $r/8$ and $r/4$.

The reduction to a single flexural mode further requires the initial bend to be symmetric and single-lobed. Bends with an antisymmetric, S-shaped component are not described by it, as observed for two of the measured configurations (Section~\ref{subsubsec:outliers}).

Both the derivation and the numerical simulations assume that the segment carries no initial axial tension, which describes a cable that has been laid down freely. An initial tension stiffens the segment against the change in sag, which raises the efficiency at a given $\Theta$ by a few percent for a realistically achievable amount of initial tension (Text S2 in Supporting Information S1). Tension is, however, difficult to establish and maintain in an unburied deployment, and the tension-free case is the relevant one for a draped cable.

The assumed ground motion and contact are idealized in the following respects. The contact points are assumed to move identically to the surrounding subsurface, which holds for a cable resting on compact soil or rock. On vegetated ground it does not: compliant material such as grass deforms under the cable, so the contact points no longer follow the bulk ground motion and the model no longer describes the coupling.
Only strain aligned with the cable axis is considered because this is what DAS predominantly senses, yet in an unburied deployment, broadside motion may couple into the axial response through the segment's dynamics. If both endpoints of a suspended segment are displaced transversely, the inertia of the cable resists the motion and stretches the segment slightly.
This contribution is second order and should be negligible in the quasi-static regime, where the segment follows the ground motion with little inertial resistance.

\subsection{Design and deployment guidelines for minimizing the sag-to-radius ratio}  

The results translate into guidelines for unburied deployments, all following from the goal of keeping the sag small relative to the radius. On the cable side, the radius is the only lever available from the specifications, which enters the strain transfer directly. The second cable property that matters is the curvature, and it is desirable that the cable retains as little curvature as possible after unrolling. This property is a consequence of the cable's construction and handling history and not a quantity that can be looked up, so it has to be established by measurement.
The Young's modulus and density do not enter the strain transfer directly. They act only through the degree of curvature that the cable retains.

On the deployment side, the suspended segment length matters most.
Pressing the cable into the ground or weighting it shortens the suspended segments and flattens the retained bend, consistent with the improvements reported for weighted and pinned surface deployments \cite{harmon_surface_2022, mjehovich_rapid_2023}, though these have not been found to help consistently \cite{forbriger_calibration_2025}. 
Because the sag cannot be predicted from cable parameters, the criterion has to be applied by measurement. Laying out a length of cable on the intended surface and measuring the characteristic midspan sags of suspended segments gives $\Theta$ directly, and the requirement is $q_{f0}/r < 1/4$ from the clamped result, or $q_{f0}/r < 1/8$ from the pinned one, which is the more conservative choice since real deployments lie between the two.
If instead the retained curvature radius $R$ is characterized, the criterion can be inverted to give the longest span a segment can bridge, $L_\mathrm{max} \approx \sqrt{2\,r\,R}$ (Equation~\ref{eq:Lmax}) for the clamped case and $\sqrt{r\,R}$ for the pinned one. This gives a target that a deployment has to meet, which indicates how flat a surface is required and how firmly the cable has to be pressed down. 

\subsection{Implications for deployment on the Moon}  

The strain transfer of a single segment does not depend directly on gravity. The efficiency depends on the sag and radius alone, so neither the relation nor the experimental validation is specific to Earth. 
Within the bending mechanism, gravity enters only through the sag. Its contribution is already small on Earth, where retained curvature dominates for spans shorter than a few meters (Equation~\ref{eq:Lstar}), and under the weaker lunar gravity it is even smaller. Gravity may nevertheless affect the coupling through the deployment method, for example through the impact of a cable dropped from a rover or the load of a wheel used to press it down, both of which scale with gravity.

Lower gravity also reduces the normal force at the contact points, and with it the friction available to resist sliding. The regolith may partly offset this: lunar regolith grains are highly angular and elongated, since they have not been rounded by weathering and transport \cite{colwell_lunar_2007}. This angularity may increase the friction coefficient between the cable and the ground.
Whether slip occurs can be estimated from the force balance of \citeA{probst_controlled_2026}. For the cable to follow the ground motion, the change of its axial force along the cable has to be supplied by friction at the contact points, which provides at most $\mu \rho A g$ per unit length, where $\mu$ is the friction coefficient between cable and ground. For a plane wave traveling along the cable with apparent velocity $c$, the strain gradient is $\partial\varepsilon/\partial x = a/c^2$, where $a$ is the ground acceleration along the cable. Slip is therefore avoided as long as
$a < \mu g \rho c^2/E$.
For lunar gravity ($g = 1.62$~m\,s$^{-2}$), friction coefficients of 0.25-0.7, apparent velocities between the regolith P-wave velocity of about 100~m\,s$^{-1}$ \cite{watkins_apollo_1972} and the 300~m\,s$^{-1}$ adopted by \citeA{zhai_assessing_2024}, and the cables of Table~\ref{tab:cables}, the threshold lies between $3\times10^{-4}$ and $6\times10^{-2}$~m\,s$^{-2}$. The strongest moonquakes recorded by the Apollo seismometers produced accelerations of about $10^{-6}$~m\,s$^{-2}$ \cite{forero_framework_2024}, at least two orders of magnitude below this threshold. Slip is therefore very unlikely for natural moonquakes recorded at regional and teleseismic distances.

For the cable itself, the weight is the largest constraint since launch mass is very costly. The radius is the property that improves the strain transfer, but it also increases the mass, so the objective is a large-radius, low-density construction. Most importantly, the cable should be designed and transported to minimize retained curvature after unrolling, since this is what sets the sag of the deployment.

\section{Conclusion}
\label{sec:conclusion}

We modeled unburied DAS cables as a sequence of beam segments suspended between discrete ground contact points, and derived an analytical solution for the axial strain transferred from the ground to the cable. The strain transfer efficiency is governed by a single dimensionless parameter, $\Theta = \frac{1}{2}\left(q_{f0}/r\right)^2$, determined by the ratio of the segment's sag to its radius. When the sag is small relative to the radius ($q_{f0}/r < 1/4$), the segment behaves like a straight rod and transfers ground strain almost completely to the fiber. As the sag grows, the segment increasingly accommodates endpoint displacement by changing its curvature instead of its arc length, so that less of the ground strain reaches the fiber as measurable axial strain. Numerical simulations validate the analytical model and laboratory measurements on 56 suspended-segment configurations follow the prediction.

The measurements show that the initial sag is dominated by the curvature the cable retains from storage and handling, which exceeds the gravitational contribution by up to several orders of magnitude. $\Theta$ is therefore a property of the cable's handling history and of the surface it is deployed on, and must be measured rather than computed. Coupling improves for cables with a larger radius, for cables that retain little curvature after unrolling, and for shorter segments between contact points.

To our knowledge, this is the first quantitative framework for ground-to-cable coupling in unburied DAS, and it offers a criterion for when a given deployment will couple well. 
The threshold of $q_{f0}/r < 1/4$ assumes contact points that clamp the cable, and tightens to $1/8$ where they do not, so the stricter value is the safer target for a deployment. Because any additional mechanism, such as frictional slip, can only reduce transfer further, $\Theta$ sets an upper bound on the achievable strain transfer. 
The coupling criterion itself applies unaffected by the lower lunar gravity. The reduced normal force at the contact points makes frictional slip more likely, but it remains unlikely for natural moonquakes.
Predicting the retained curvature and contact spacing that set $\Theta$ in the field is the next step toward a complete picture of unburied DAS coupling and a basis for optimizing deployments on both Earth and the Moon. 

\section*{Open Research Section}
The data and processing scripts used in this work are openly available on the ETH Zurich Research Collection via https://doi.org/10.3929/ethz-c-000806579 \cite{probst_supporting_2026}.

\section*{Conflict of Interest disclosure}
The authors declare there are no conflicts of interest for this manuscript.

\acknowledgments
We thank Henrik Thomsen and Christoph Bärlocher for their support with the laboratory experiments. 
SP acknowledges funding from the Swiss National Science Foundation, Grant 200021\_212064/1.
CMD was supported by the Los Alamos National Laboratory (LANL) through its Center for Space and Earth Science (CSES). CSES is funded by LANL’s Laboratory Directed Research and Development (LDRD) program under project number 20240477CR.
This manuscript has been assigned the number LA-UR-26-26682 by Los Alamos National Laboratory.

\bibliography{references}

\end{document}

\title{Supporting Information for ``Ground-to-Cable Strain Transfer in Unburied DAS on Earth and the Moon''}

\authors{S. Probst\affil{1}, O. Cornelius\affil{1}, M. Serra Garcia\affil{2}, J. O. A. Robertsson\affil{1}, C. M. Donahue\affil{3}}

\vspace{0.5cm} % added by SP

\affiliation{1}{Department of Earth and Planetary Sciences, ETH Zurich, Zurich, Switzerland}
\affiliation{2}{AMOLF, Amsterdam, The Netherlands}
\affiliation{3}{Earth and Environmental Sciences, Los Alamos National Laboratory, NM, USA}

\begin{article}

\noindent\textbf{Contents of this file}
\begin{enumerate}
\item Texts S1 to S4
\item Figures S1 to S8
\end{enumerate}

\medskip
\noindent\textbf{Introduction}

This supporting information provides additional details on the numerical model and the laboratory experiments described in the main text. Text S1 validates the analytical prediction of gravity-induced sag against the numerical model. Text S2 assesses numerically how an initial axial tension in the cable segment changes the strain transfer efficiency. Text S3 reports the strain amplitudes imposed in the laboratory, the numerically predicted asymmetry between extension and compression at these amplitudes, and the asymmetry estimated from the measurements. Text S4 provides additional details on the dynamic response of the cable segments, including all detected resonances, the transverse motion of the shaker head, and the scaling of the first bending mode with span. The numerical simulations and the laboratory data processing are described in Sections 3.1 and 3.3 of the main text. % MAINREF

\bigskip
\noindent\textbf{Text S1. Gravity-Induced Sag Magnitude}
\label{app:gravity_sag}

The theoretical model predicts that the magnitude of the sag induced by gravity acting on an initially straight beam segment is $q_{f0}=\rho A g L^4/(4 \pi^4 E I)$ (Equation 16 of the main text). % MAINREF
We validate this theoretical result through comparison with the numerical model. Figure~\ref{fig:app_gravity_sag} shows the numerically modeled sag against the theoretical prediction. The modeled values are the same as given in Section 3.1 of the main text, but only cases with no retained curvature and non-zero gravity were modeled. The numerical results validate the theoretical prediction for the investigated range. % MAINREF

\begin{sifigure}
    \noindent\includegraphics[width=0.5\textwidth]{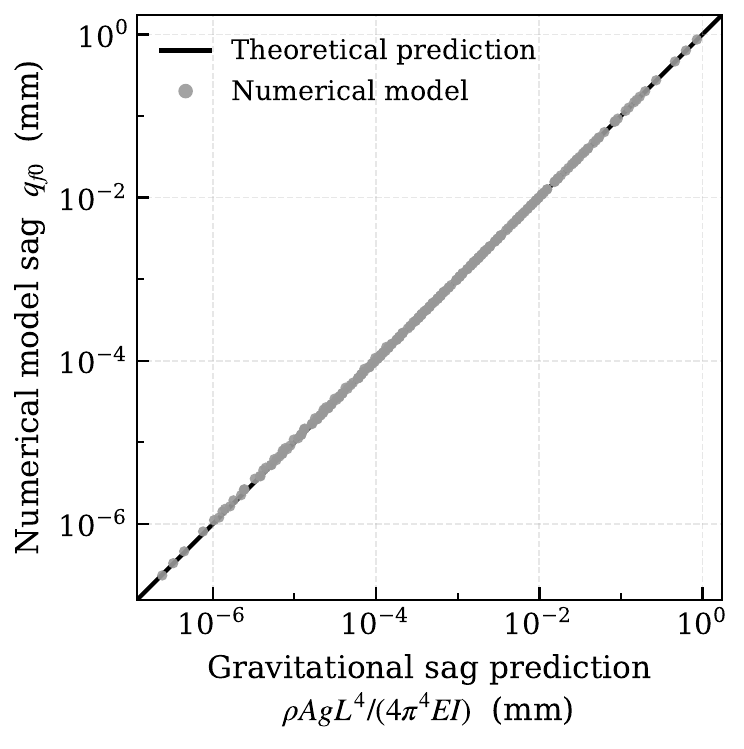}
    \caption{Magnitude of gravity-induced sag. Comparison of theoretical prediction against numerical model results.}
    \label{fig:app_gravity_sag}
\end{sifigure}

\bigskip
\noindent\textbf{Text S2. Initial Axial Tension}
\label{app:axial_tension}

The derivation in Section 2.2 of the main text and the simulations in the main text assume that the segment carries no axial tension at rest. A cable that is pulled taut during deployment would violate this assumption. Here we assess numerically how an initial tension changes the strain transfer efficiency. % MAINREF

Each run starts from the equilibrium shape of a segment whose load-free configuration is a circular arc of radius $R$, as in Section 3.1 of the main text. The endpoints are then moved apart statically by a distance $D$, which partially straightens the segment and builds up an axial tension $T$. The strain transfer efficiency is evaluated about this tensioned equilibrium by imposing a further endpoint displacement corresponding to 1~$\mu\varepsilon$, and $\Theta$ is evaluated from the sag of the tensioned segment. Gravity is omitted, so the density does not enter. % MAINREF
The span was set to L = 5, 10 and 20~cm, the radius to r = 0.25, 0.5, 0.75 and 1~mm, and the retained curvature radius to R = 0.25, 0.5, 1, 2.5 and 5~m, with a Young's modulus of E = 5.48~GPa. Each combination was run at the imposed pre-strains $D/L = 0$, $10^{-4}$, $10^{-3}$ and $10^{-2}$, resulting in 240 runs. The imposed pre-strains produce tensions of up to 172~N.

The results are shown in Figure~\ref{fig:app_tension}. The untensioned runs lie on the analytical curve over the full range of $\Theta$. Tension improves the strain transfer in two ways. First, it removes part of the retained bend, which reduces the sag and moves the segment along the curve toward higher efficiency. Second, a tensioned segment transfers more strain than an untensioned segment with the same sag, because the tension resists the change in sag through which the segment accommodates the endpoint displacement.

Tension would therefore improve the coupling of a draped cable, but it is difficult to establish and maintain in the field. Without anchors, the tension has to be provided by friction along the cable. Tension also lifts the cable off depressions in the surface, so that it touches down less often and its suspended segments become longer.

\begin{sifigure}
    \noindent\includegraphics[width=\textwidth]{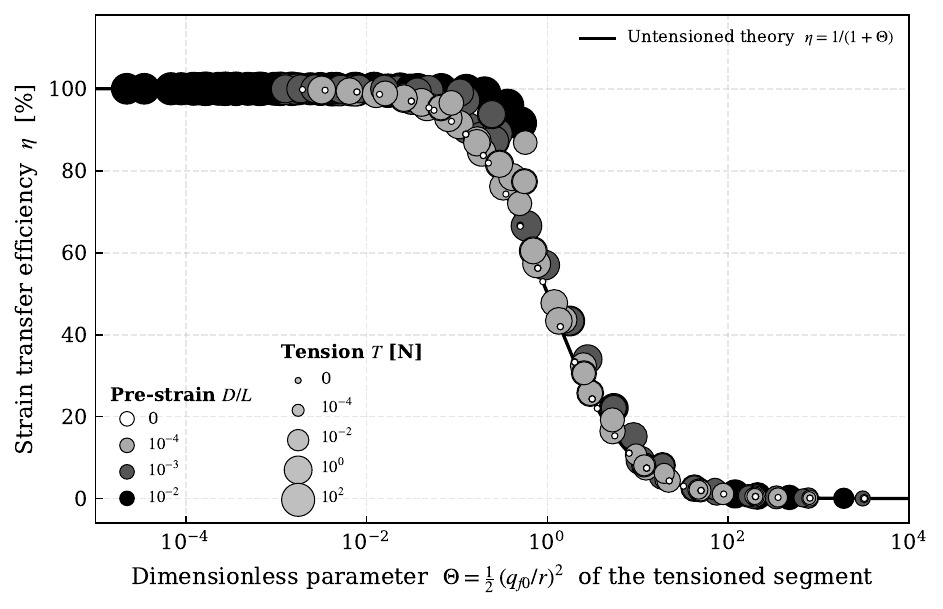}
    \caption{\textbf{A tensioned segment transfers strain at least as efficiently as an untensioned segment with the same sag.}
    Strain transfer efficiency $\eta$ against $\Theta$ evaluated from the sag of the tensioned segment, for 240 simulated segments with retained curvature and no gravity. Marker shading gives the pre-strain $D/L$ imposed before the efficiency is evaluated, and marker size the resulting axial tension $T$. White markers are untensioned ($D = 0$). The solid line is the untensioned analytical prediction (Equation 12 of the main text).} % MAINREF
    \label{fig:app_tension}
\end{sifigure}

\bigskip
\noindent\textbf{Text S3. Extension-Compression Asymmetry at the Applied Amplitudes}
\label{app:linearity}

The model predicts that a suspended segment responds with a different efficiency in extension than in compression once the imposed displacement is no longer small compared with the length stored in its initial bend (Section 4.2 of the main text). This text gives the laboratory strain amplitudes (Text S3.1), the numerical prediction of the asymmetry for the measured configurations and the efficiency values when processing the numerical results with the same approach as the experimental data (Text S3.2), and the asymmetry observed in the experimental data (Text S3.3). % MAINREF

\medskip
\noindent\textbf{\textit{S3.1. Laboratory Strain Amplitudes}}
\label{app:imposed_strains}

Figure~\ref{fig:app_imposed_strains} shows the amplitudes of the strain imposed at the segment endpoints for each configuration. These amplitudes were calculated from the band-mean ($<$ 25 Hz) endpoint displacements.

\medskip
\noindent\textbf{\textit{S3.2. Numerical Prediction}}

The prediction comes from the second set of simulations described in Section 3.1 of the main text. Each of the 56 configurations is reproduced with its measured sag and cross-section and loaded statically at the strain amplitude imposed on that configuration, once in extension and once in compression, giving $\eta_\mathrm{ext}$ and $\eta_\mathrm{comp}$ at that configuration's own drive amplitude. The resulting asymmetry is shown in Figure~\ref{fig:app_num_asymmetry}a. % MAINREF
The results of the numerical model were then processed in a similar manner as the experimental data, to assess whether the asymmetry of extension and compression affects the estimated efficiencies, or whether the efficiencies averaged over extension and compression cycles fall on the same curve.
The strain was projected along the static segment direction as for the experimental data and then averaged between compression and extension.
The results are shown in Figure~\ref{fig:app_num_asymmetry}b. Only three configurations deviate substantially from the small-amplitude prediction. One of these is the configuration which was excluded from the results, because none of its frequency bins exceeded the coherence threshold of 0.7, the other two correspond to discussed outlier cases.

\medskip
\noindent\textbf{\textit{S3.3. Assessing Asymmetry in Experimental Data}}

The estimator of Equation 22 of the main text is evaluated at the drive frequency, which removes the asymmetry from the data, since the asymmetry has energy at DC and at even multiples of the drive frequency. Therefore, a different evaluation is necessary to assess whether the experimental data shows a compression-extension asymmetry. We calculate the elongations sample by sample and apply no band-pass around the drive frequency, since such a filter would remove the asymmetry which we want to quantify. The three velocity components at every scan point are band-passed over the full record between 3 and 75~Hz and integrated to displacement as described in Section 3.3 of the main text. The frequencies below 3 Hz are removed to avoid low-frequency noise contaminating the signal with a signal drift. For each time sample the cable elongation is evaluated as the exact change in three-dimensional arc length, % MAINREF
summed over the sections between adjacent scan points.
Unlike the estimator of Section 3.3 of the main text, the exact arc-length change makes no small-slope approximation. % MAINREF

To assess the asymmetry, the strain transfer efficiencies are then determined separately for compression and extension time samples. This is done by fitting a regression with a different slope for positive and negative strains,
\begin{equation}
\label{eq:rectfit}
\delta q_l = c + \bar{\eta}\, \delta L + \frac{\Delta \eta}{2} \left| \delta L \right| ,
\qquad
\bar{\eta} = \frac{\eta_\mathrm{ext} + \eta_\mathrm{comp}}{2} ,
\qquad
\Delta \eta = \eta_\mathrm{ext} - \eta_\mathrm{comp} ,
\end{equation}
solved by least squares over all samples in the window, from which $\eta_\mathrm{ext} = \bar{\eta} + \Delta \eta/2$ and $\eta_\mathrm{comp} = \bar{\eta} - \Delta \eta/2$ follow directly. The rectified basis function $\left| \delta L \right|$ is the only regressor that distinguishes the two half cycles, so the asymmetry is carried by a single coefficient. The constant $c$ is fitted rather than set to zero. Fitting the two half cycles separately with lines forced through the origin, which is the more obvious construction, makes any residual low-frequency shift in the strain values tilt the two slopes in opposite directions and generate an apparent asymmetry from an offset alone.

Figure~\ref{fig:asymmetry_model_vs_measurement} plots the measured and predicted asymmetry.
The individual values scatter around the prediction. Two reasons account for this. The estimator works on the full record rather than at a single frequency and has no averaging over repeated realizations available to it, so it inherits the full measurement noise, in contrast to the Welch-averaged transfer function of Section 3.3 of the main text. In addition, the configurations for which the asymmetry is predicted to be largest are largely those with the least reliable efficiencies in the first place, in particular the thinnest cable C7, whose measurements show low coherence and large variability across the quasi-static band (Section 4.2 of the main text). The comparison therefore supports the presence of the predicted effect and its dependence on $\Theta$, but does not constitute a quantitative test of its magnitude. % MAINREF

\begin{sifigure}
    \noindent\includegraphics[width=\textwidth]{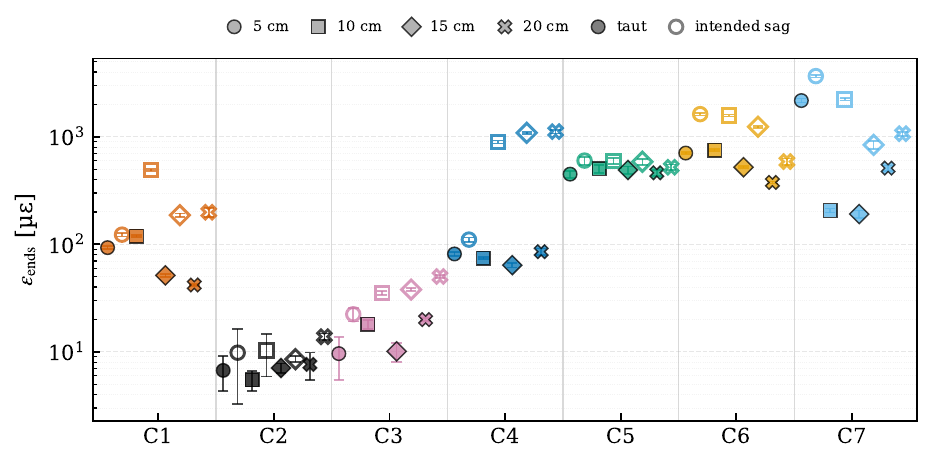}
    \caption{Imposed strain per configuration. $\varepsilon_{ends}$ is calculated as $\delta L/L$.}
    \label{fig:app_imposed_strains}
\end{sifigure}

\begin{sifigure}
    \noindent\includegraphics[width=\textwidth]{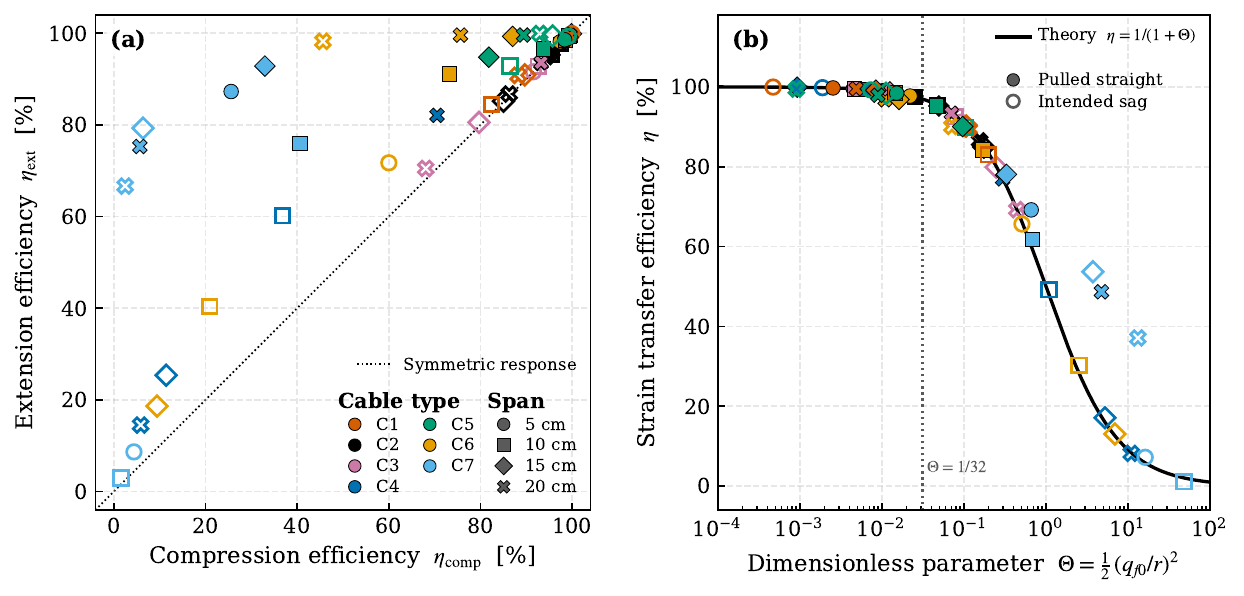}
    \caption{Numerically evaluated asymmetry between extensional and compressional strain at the configuration-specific experimental strain amplitudes. Filled markers indicate taut configurations while outline-only markers indicate configurations with intended sag. \textbf{(a)} Extensional efficiency vs compressional efficiency. \textbf{(b)} Resulting averaged strain transfer efficiencies compared to the small-amplitude theoretical prediction.}
    \label{fig:app_num_asymmetry}
\end{sifigure}

\begin{sifigure}
    \noindent\includegraphics[width=\textwidth]{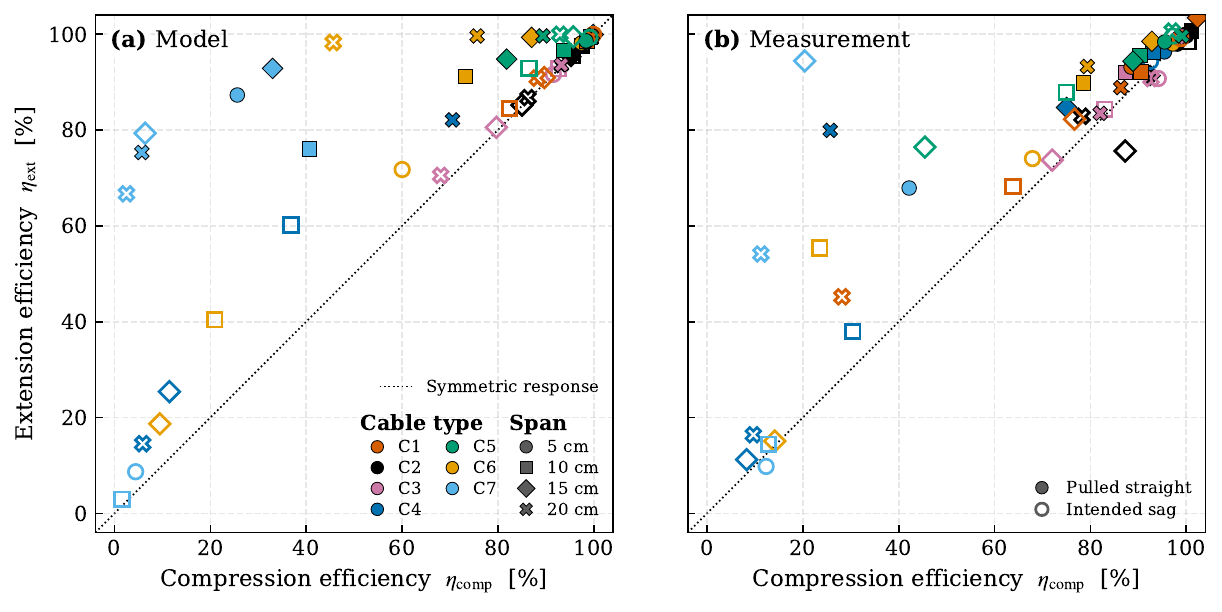}
    \caption{\textbf{Model and measurement both respond more efficiently in extension than in compression, but they do not agree configuration by configuration.}
    \textbf{(a)} Extension against compression efficiency of the numerical model. Each laboratory configuration is reproduced with its measured sag and cross-section and loaded statically at its measured endpoint strain, once in extension and once in compression.
    \textbf{(b)} The same quantities obtained from the measurements with the sample-wise estimator.
    Both panels show the same trend: all modeled and most measured configurations (46 of 56) lie above the line, transferring strain more efficiently in extension. Individual configurations, however, do not coincide between model and measurement.
    Four measured configurations fall outside the axes. Three of them, C7 at the 10, 15 and 20\,cm spans without intended sag, are the recordings with the lowest coherence in the data set (mean $\gamma^2$ of 0.14, 0.76 and 0.15, against a median of 0.995). Unlike the transfer-function estimate of Section 3.3 of the main text, the sample-wise estimator does not average over repeated realizations, and it returns no meaningful values for these noisy recordings.
    Configurations identified as outliers in Figure 4 of the main text depart from the model here as well. This is expected: where the efficiency averaged over the two half cycles does not follow the prediction, the extension and compression efficiencies cannot be expected to do so individually.} % MAINREF
    \label{fig:asymmetry_model_vs_measurement}
\end{sifigure}

\bigskip
\noindent\textbf{Text S4. Dynamic Response}

\medskip
\noindent\textbf{\textit{S4.1. Cable Resonance Frequencies}}
\label{app:detected_resonances}

All detected resonances are shown in Figure~\ref{fig:resonance_overview}.

\medskip
\noindent\textbf{\textit{S4.2. Transverse Input of the Shaker}}
\label{app:shaker_direction}

The shaker is intended to drive the cable along its axis, but at certain frequencies its head also moves transversely. Figure~\ref{fig:app_shaker_direction} shows this for one configuration, together with the resulting transfer functions. The transverse-to-axial ratio $R_\mathrm{sh}$ (Equation~\ref{eq:Rsh}) is calculated as
\begin{equation}
\label{eq:Rsh}
R_\mathrm{sh}(f) = \sqrt{\frac{S_{v_y v_y}(f) + S_{v_z v_z}(f)}{S_{v_x v_x}(f)}} ,
\end{equation}
with $x$ along the drive axis. Where $R_{sh}(f)$ rises above one, the midspan transfer function shows a peak, and the strain transfer function shows the corresponding dip and overshoot, although no bending mode of the segment is predicted there. The same band appears in the recording of the bare shaker, which identifies it as a property of the shaker rather than of the cable.

\medskip
\noindent\textbf{\textit{S4.3. First Bending Mode Against Span}}
\label{app:span_scaling}

The remaining peaks are compared with the fundamental frequency of a straight Euler-Bernoulli beam,
\begin{equation}
\label{eq:f1_bc}
    f_1 = \alpha_1^2/(2\pi L^2)\sqrt{EI/(\rho A)},
\end{equation}
evaluated for clamped ($\alpha_1 = 4.730$, Equation 24 of the main text) and for pinned ($\alpha_1 = \pi$) endpoints, for the parameters of each configuration. % MAINREF
Because Equation~\ref{eq:f1_bc} scales as $L^{-2}$ for any fixed boundary condition and any modulus, the first bending mode of each cable is additionally fitted as $f_1 = C L^{-p}$ over the spans of 10, 15 and 20~cm, separately for the taut and the intended-sag configurations. Here the first bending mode is the lowest peak that is not classified as driven by transverse shaker motion.

For a beam with a boundary condition that does not change with span, the fundamental frequency scales as $L^{-2}$, independently of the modulus (Equation~\ref{eq:f1_bc}). A lower modulus shifts the frequencies down without changing this scaling, whereas a support whose influence depends on the span changes the exponent. Figure~\ref{fig:app_span_scaling} shows the first bending mode against the span for each cable, together with the fit $f_1 = C L^{-p}$ and the range between the pinned and clamped predictions.

The first modes of most cables lie below or at the lower end of the predicted range. 

\begin{sifigure}
    \noindent\includegraphics[width=\textwidth]{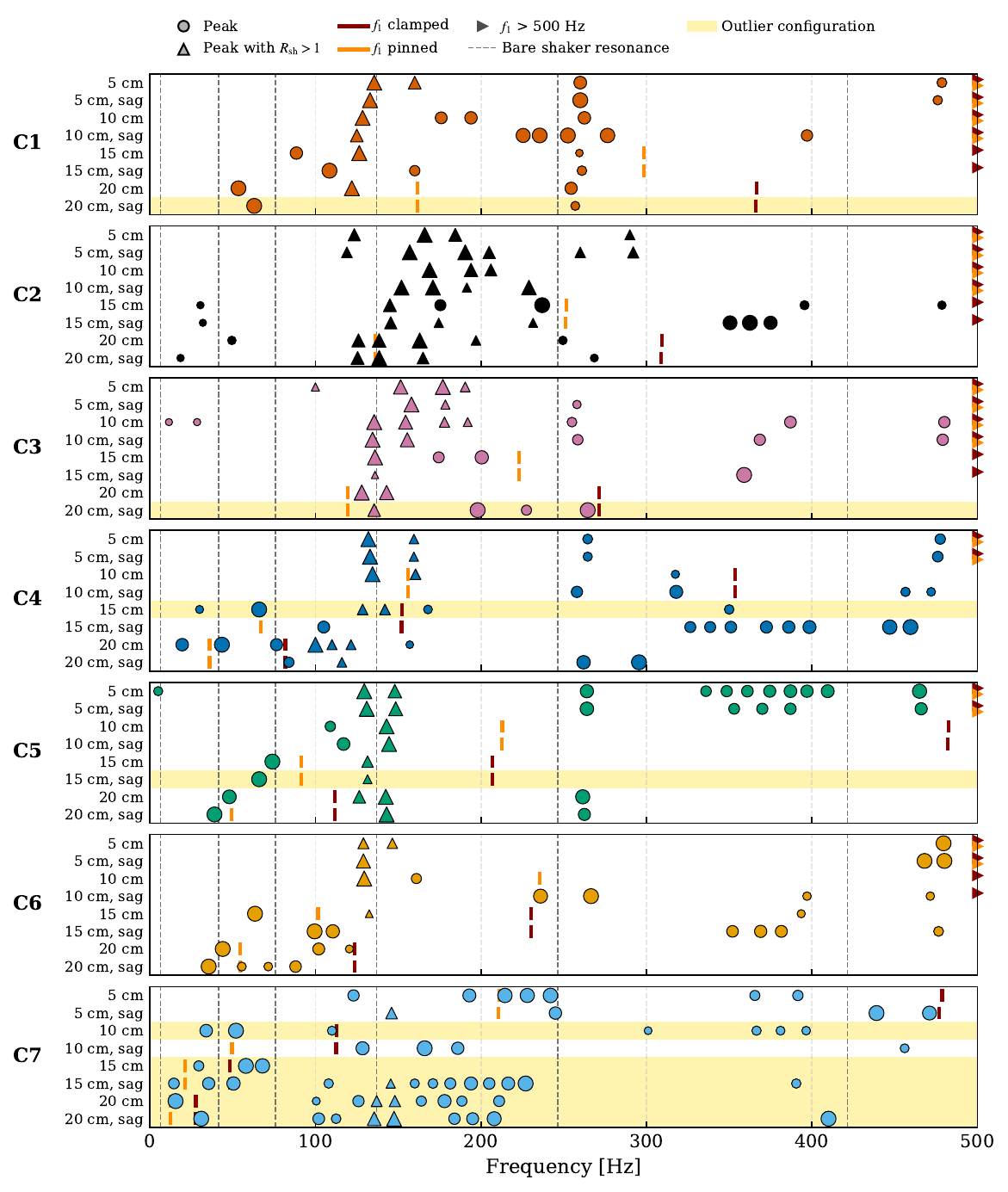}% reduced so figure + caption fit on one page
    \caption{\textbf{Peaks of the midspan transfer function fall into a band driven by transverse shaker motion and a set of lower bending modes that lie below the predicted fundamental frequency.}
    Each panel shows one cable, with one row per configuration (span, and taut or intended sag). Markers are the identified peaks, with triangles for peaks where the shaker head moves more transversely than axially ($R_\mathrm{sh} > 1$) and circles otherwise; marker size scales with the normalized midspan amplitude. Vertical bars mark the fundamental frequency predicted for clamped (dark red) and pinned (orange) endpoints (Equation~\ref{eq:f1_bc}); arrowheads at the right edge indicate predictions above 500~Hz. Dashed vertical lines are resonances of the bare shaker. Configurations identified as outliers in Section 4.2 of the main text are not shown.} % MAINREF
    \label{fig:resonance_overview}
\end{sifigure}

\begin{sifigure}
    \noindent\includegraphics[width=\textwidth]{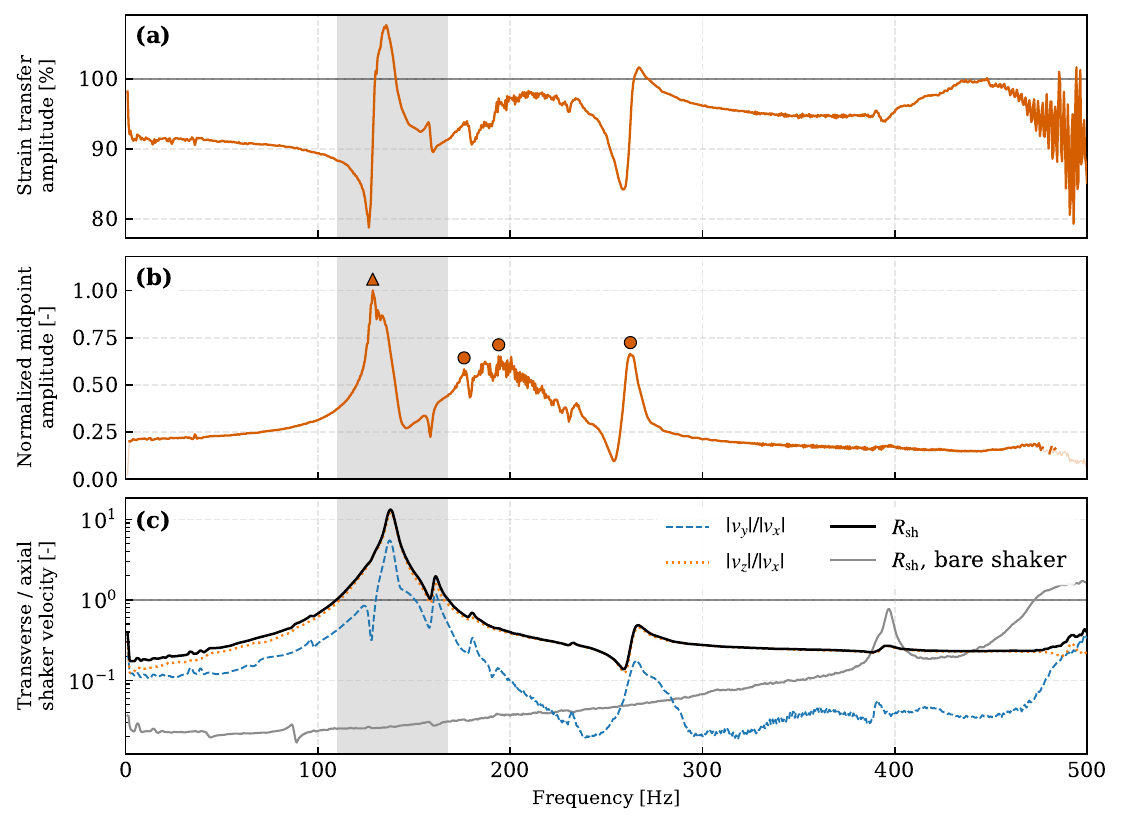}
    \caption{\textbf{The shaker head moves transversely at specific frequencies, producing peaks in the midspan transfer function that are not bending modes of the segment.}
    Configuration [C1, 10~cm, taut].
    \textbf{(a)} Strain transfer amplitude $|H_\eta|$.
    \textbf{(b)} Normalized midspan amplitude $|H_\mathrm{mid}|$ with the identified peaks.
    \textbf{(c)} Ratios of transverse to axial velocity amplitude of the shaker head, $|v_y|/|v_x|$, $|v_z|/|v_x|$ and $R_\mathrm{sh}$, recorded with the cable attached. The horizontal line marks a ratio of one.}
    \label{fig:app_shaker_direction}
\end{sifigure}

\begin{sifigure}
    \noindent\includegraphics[width=\textwidth]{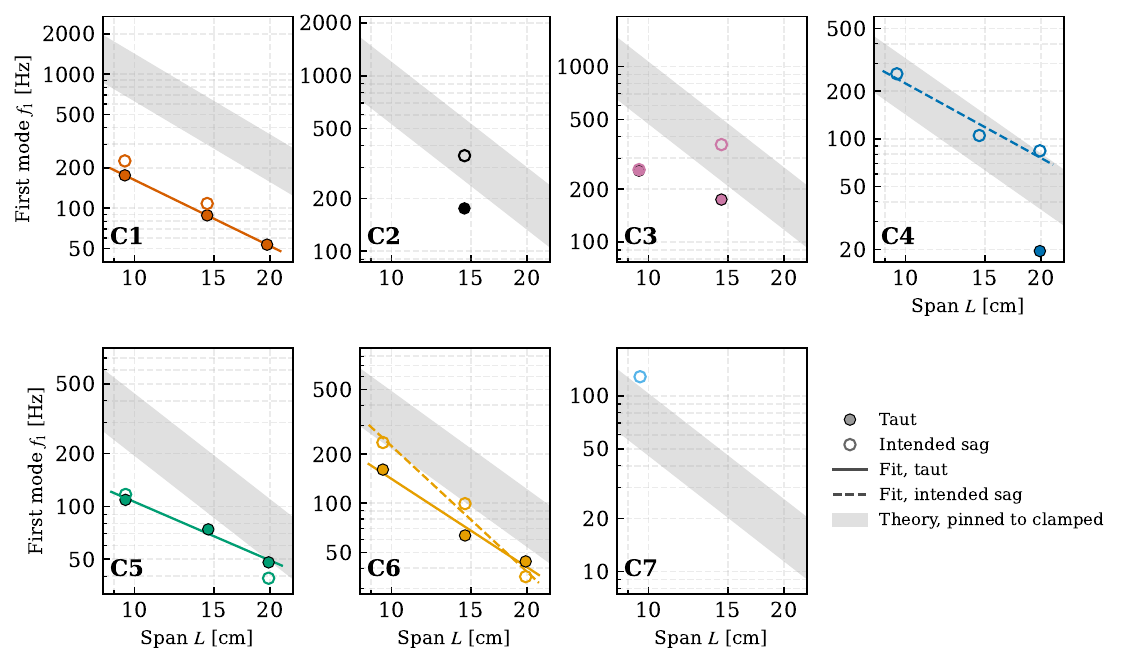}
    \caption{\textbf{The first bending mode decreases with span, but lies below the prediction and decreases more slowly than $L^{-2}$.}
    Frequency of the first bending mode against the clamp-to-clamp span, one panel per cable. Filled markers are taut configurations, open markers configurations with intended sag. Lines are fits $f_1 = C L^{-p}$. The grey band spans the predictions for pinned (lower edge) and clamped (upper edge) endpoints with the tensile modulus (Equation~\ref{eq:f1_bc}). Configurations identified as outliers in Section 4.2 of the main text are not shown.} % MAINREF
    \label{fig:app_span_scaling}
\end{sifigure}

\end{article}
\clearpage